\documentclass[11pt,leqno,a4paper]{book}

\usepackage{amsfonts,amsmath,amssymb,amsthm,bbm}

\usepackage{psfrag}
\usepackage[pdftex]{graphicx}
\usepackage{color}                    

\usepackage[breaklinks=true,bookmarks=false,linkbordercolor={0 0 1}]{hyperref}                 

\newcommand{\HRule}{\rule{\linewidth}{0.5mm}}

\newcommand{\Ref}[1]{(\ref{#1})}
\newcommand{\tr}{\mathrm{tr}}

\def\be{\begin{equation}}
\def\ee{\end{equation}}
\def\bes{\begin{eqnarray}}
\def\ees{\end{eqnarray}}
\def\nn{\nonumber}

\newcommand{\N}{\mathbb{N}}
\newcommand{\C}{\mathbb{C}}

\newcommand{\R}{\mathbb{R}}

\newcommand{\su}{\mathfrak{su}}
\renewcommand{\sl}{\mathfrak{sl}}
\newcommand{\ag}{\mathfrak{g}}
\newcommand{\SU}{\mathrm{SU}}
\newcommand{\SL}{\mathrm{SL}}
\newcommand{\U}{\mathrm{U}}
\newcommand{\SO}{\mathrm{SO}}
\newcommand{\ISO}{\mathrm{ISO}}

\newcommand{\Spin}{\mathrm{Spin}}

\def\f{\frac}
\def\pp{\partial}
\def\w{\wedge}
\def\la{\langle}
\def\ra{\rangle}
\def\arr{\rightarrow}

\def\om{\omega}
\def\eps{\epsilon}
\def\vphi{\varphi}

\def\tl{\widetilde}
\def\wh{\widehat}
\def\tpsi{\tl{\psi}}

\newcommand{\vJ}{\vec{J}}
\newcommand{\vK}{\vec{K}}

\def\mn{{\mu\nu}}
\def\bv{{\bar{v}}}
\def\bl{{\bar{l}}}

\def\cA{{\cal A}}
\def\cM{{\cal M}}
\def\cG{{\cal G}}
\def\cV{{\cal V}}
\def\cR{{\cal R}}
\def\cO{{\cal O}}
\def\cI{{\cal I}}
\def\cH{{\cal H}}
\def\cP{{\cal P}}
\def\cS{{\cal S}}

\def\cK{{\cal K}}

\begin{document}


\begin{titlepage}

\vspace*{-15mm}
\begin{minipage}{60mm}
\begin{flushleft}
\includegraphics[height=15mm]{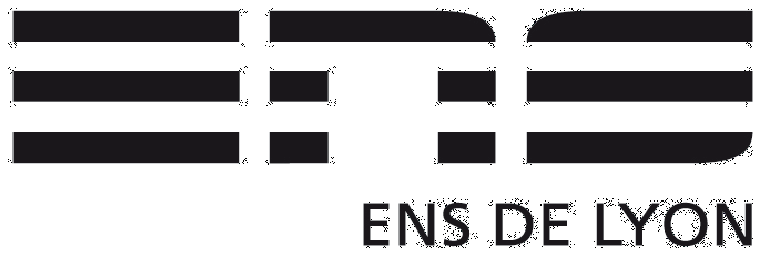}
\end{flushleft}
\end{minipage}
\begin{minipage}{60mm}
\begin{flushright}
\includegraphics[height=25mm]{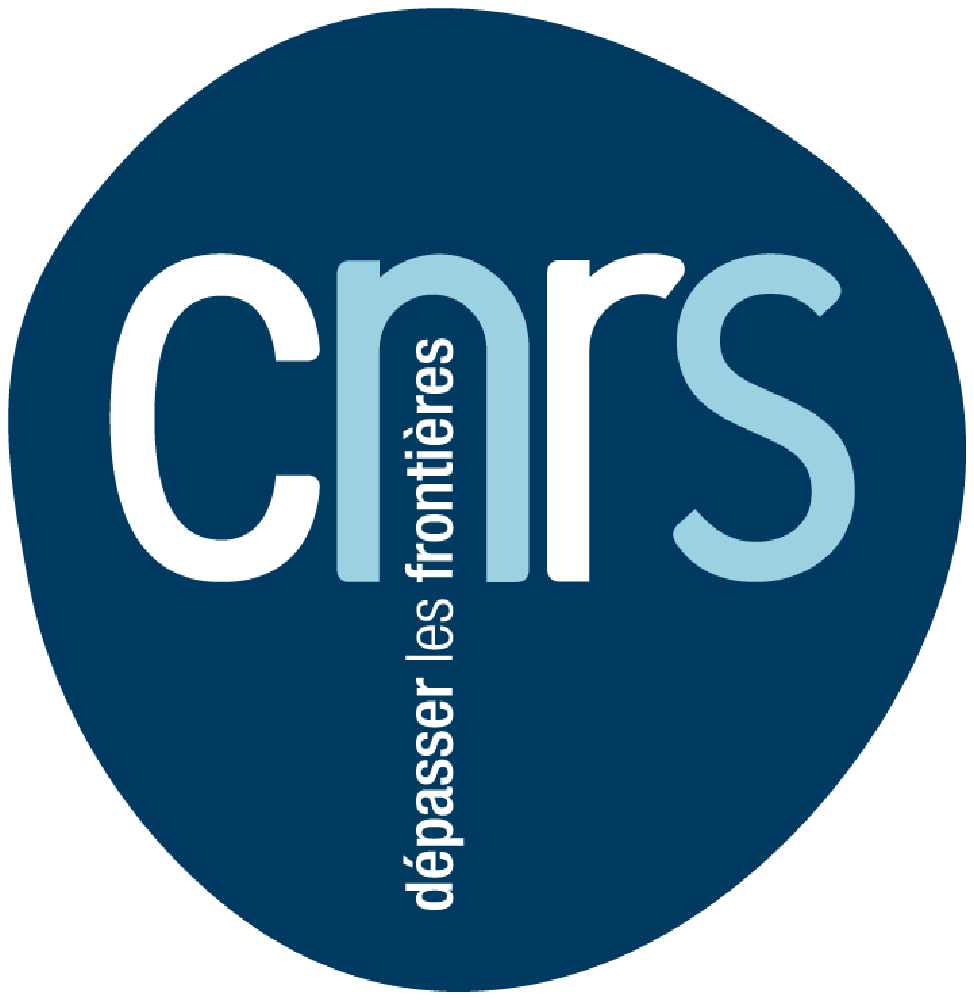}
\end{flushright}
\end{minipage}

\vspace{35mm}

\begin{center}

\textsc{\LARGE Ecole Normale Sup\'erieure de Lyon}\\[2mm]
\textsc{\Large Laboratoire de Physique - CNRS UMR 5672}\\[20mm]

\textsc{\LARGE Th\`ese d'Habilitation \`a Diriger des Recherches}\\[15mm]

\HRule \\[0.4cm]
{ \Huge \bfseries The Spinfoam Framework for \\[2mm] Quantum Gravity}\\[0.4cm]

\HRule \\[1.5cm]

\begin{center}
{\LARGE Etera \textsc{Livine}}
\end{center}

\vspace{20mm}

{\large October 2010}

\end{center}




\end{titlepage}

 \addcontentsline{toc}{chapter}{Contents}
\pagenumbering{roman}


\tableofcontents


\pagestyle{headings}
\pagenumbering{arabic}


\part{Spinfoams and Quantum Gravity}

\chapter{Introduction: Searching for Quantum Gravity}

Quantum gravity is the search for a physical theory describing the gravitational interaction at all scales of length and energy. It is one of the biggest problems and challenges of theoretical physics, which we have inherited from the twentieth century (see e.g. \cite{incompleterev} for a short historical review of the research on quantum gravity). This theory must be valid in cosmological settings as well as for microscopic phenomena at the shortest scale, and it should reunify quantum mechanics, which describes the microscopic laws of our universe and the propagation of particles, with general relativity, which describes the dynamics of the gravitational field and the coupling of matter to gravity.

On the one hand, quantum mechanics provides a completely probabilistic description of matter evolving in a fixed rigid space-time. All the information is encoded in the wave-function, which defines the probability amplitude of possible events. Then the theory defines the equations of motion that this wave-function must satisfy. The original formulation has later evolved into quantum field theory, where the wave-function itself becomes the fundamental degree of freedom. This is the framework for the standard model of particle physics. Up to now, this theory describes perfectly all microscopic phenomena, the scattering of particles and their interactions.

On the other hand, general relativity describes the gravitational interaction, which dominates all other (known) interactions at large distances and thus defines the large scale structure of the universe. Unlike the other interactions, gravity is encoded in the fabric of space-time itself: matter and energy interact with the space-time and curve it and deform it. The fundamental degree of freedom is the metric of the space-time manifold and the theory defines its dynamics through the Einstein equation, which relates the space-time curvature to the energy content evolving in that space-time.

The goal of quantum gravity is to unify these two frameworks and write a consistent physical theory which can describe gravitational interactions and the dynamics of the metric  itself at the microscopic quantum scales. This unification needs to be both conceptual and mathematical, since indeed these two theories are both based on very different principles and using different mathematics to define the dynamics and evolution of matter and space-time.

\medskip

The simplest way to see that we are expecting new physics from merging quantum field theory with general relativity is to combine the fundamental constants of these theories: the speed of light $c$, the Planck constant $\hbar$ which define a unit of action, and finally Newton's constant $G_N$ which defines the strength of the gravitational interaction. Indeed, combining these constants, we see a new scale emerging: the Planck length $l_P=\sqrt{\hbar G_N/c^3}\sim 1.62\,10^{-35}\,m$ or equivalently the Planck mass $m_P=\sqrt{\hbar c/G_N}$. At this microscopic scale $l_P$, we expect large quantum fluctuations of the gravitational field, which would affect the microscopic structure of space-time. To understand the physical relevance of the Planck scale, let us imagine attempting to measure a given distance $l$ using a laser beam. To this purpose, we need a laser beam with a wavelength $\lambda$ smaller than $l$. Then assuming that quantum field theory is valid at this scale, the energy carried by the laser beam is $E=\hbar\om=2\pi\hbar c/\lambda$. As $\lambda$ gets smaller, this energy increases. The more it increases, the more it deforms the surrounding space-time. Assuming that the laws of general relativity holds at this scale, this energy eventually reaches the Schwarzschild bound $E_{max}=\lambda c^4/2G_N$ when the laser beam would carry enough energy to collapse into a microscopic black hole. This scale is given by $E=E_{max}$ which leads to $\lambda=l_P\,\sqrt{4\pi}$. This means that we can not measure events occurring on smaller distances than $l_P$. This means that, assuming that the laws of both quantum mechanics and general relativity are valid at those scales, we should give up the notion of fundamental space-time points and our implicit assumption than space-time is a continuous manifold.
Thus this simple argument shows that the laws of quantum mechanics and general relativity can not be valid at those scales and that their principles need to be modified in order to unify these theories together.

Beyond this simplest reason, we can give several more precise arguments and motivations to look for a theory of quantum gravity:
\begin{itemize}

\item {\bf Singularities and the Incompleteness of General Relativity}: General relativity leads naturally and inevitably to singularities, such as black holes or the big bang singularity, where some components of the gravitational field diverge. These infinities mean that general relativity ceases to be a valid theory of gravity in these regimes and we need to look for a more general theory which could describe realistically the physics in those extreme situations. These singularities happen when the concentration of matter and energy is too high and when, as a consequence, the matter will collapse onto itself ultimately leading to regions of space-time with infinite curvature. It is believed that quantum effects should become extremely large in those situations, with huge concentrations of matter on microscopic scales, and that therefore properly describing the quantum fluctuations of the gravitational field will allow to resolve those singularities.

\item {\bf Conceptual Incompleteness of the Quantum Theory}: Quantum mechanics is very rigorously constructed and very well tested experimentally. It nevertheless gives the impression to lack a proper explanation, which would give the reason for all this mathematical apparatus with Hilbert spaces and quantum states. Moreover, there is still the issue of the collapse of the wave-function and the related question of the separation classical/quantum and observer/system. Although there is already a lot work done on the interpretation of quantum mechanics which tries to address those issues, it is also believed that a deeper theory such as quantum gravity, which would apply the laws of quantum mechanics not only to matter but also directly to the space-time itself, should be based on deeper principles and would help to shed lights on the true physical meaning of the quantum theory.

\item {\bf The Matter/Geometry Equivalence}: The Einstein equations for classical general relativity equate the curvature tensor for space-time with the stress-energy tensor describing its matter content. This means that matter deforms the space-time geometry accordingly to its configuration. It also means that if matter is quantized and follows the laws of quantum mechanics then the metric and geometry for space-time, as the other side of the equation, should also be quantized and be subject to the laws of quantum mechanics.

\item {\bf Quantizing Gravitational Waves}: Looking at small perturbations of the metric around the flat Minkowski space-time, we find that the gravitational field has its own degrees of freedom that can propagate and fluctuate independently of the matter source. These are given by the gravitational waves, which are classical solutions of the Einstein equations for general relativity in vacuum. Having in mind the other interactions (electromagnetism and the nuclear interactions), classical wave solutions should become (gauge) particles at the quantum level. Therefore we expect that general relativity should be quantized into a theory of quantum gravity which would describe, in a low energy regime, a physical (spin-2) particle corresponding to the classical gravity waves and its interactions with the other particles. This ``graviton" would carry the gravitational interaction the same way that the photon carries the electromagnetic force.

\item {\bf Why Not?}: General relativity is a field theory with local degrees of freedom. The field is the metric, which defines the space-time geometry. This field is dynamical, it evolves with the matter content of space-time but also has some proper degrees of freedom that can propagate independently. As a field theory with dynamical degrees of freedom, it is natural to apply to it the laws of quantum mechanics and quantize it by looking at its path integral and summing on all possible fluctuations of the gravitational field. We would need to give a good reason not to quantize it.

\item {\bf The Mysterious Black Hole Entropy}: Black holes are the most simple non-trivial, and thus the most studied, objects in general relativity. They describe the ultimate step of the gravitational evolution when matter collapses into a singularity hidden behind a horizon which blocks anything from escaping the black hole. These black holes configuration are extremely smooth since everything is hidden behind the horizon: their classical degrees of freedom are very few- their mass, charges and angular momenta. It was nevertheless found that these objects radiate at a certain temperature and that, as a consequence, they carry a non-trivial entropy. This resulting entropy does not match the very small number of classical degrees of freedom associated to black holes. It is thought that this entropy must come from non-trivial degrees of freedom of quantum gravity associated to black holes configuration. This actually provides a first non-trivial test for quantum gravity models, which need to predict the correct black hole entropy and temperature.

\end{itemize}
These are only some of the motivations to look for a theory of quantum gravity and we also hope that quantum gravity will address all of those issues. There are also other motivations, especially in the context of cosmology, but it is not always obvious that a consistent theory of quantum gravity would necessarily help solving all those problems.

\medskip

Now that we have good reasons to study quantum gravity, we simply need to write a consistent theory for it and then test it experimentally. This is actually not as straightforward as it sounds. Here are a few of the obstacles:
\begin{itemize}

\item {\bf Adapt Quantum Mechanics to a Fluctuating Space-Time}: Quantum mechanics and quantum field theory were originally formulated on a fixed flat space-time. Now, we need to adapt this framework to accommodate a curved and fluctuating space-time. Quantum field theory has already been developed on a curved background and already has a lot of new subtleties compared to standard quantum mechanics. Now, we have to go one big step more and adapt the formalism to a fluctuating space-time geometry. Furthermore, we have to do this while respecting the diffeomorphism invariance of general relativity, which is its fundamental symmetry. This diffeomorphism invariance implies some kind of background independence of the theory, which leads to think that we now have to formulate quantum mechanics not on a fluctuating space-time background but totally without referring to any background. While a lot of research has already been achieved on this topic, it seems that much is still to be understood.

\item {\bf Failure of Quantum Field Theory}: One of the motivation for quantum gravity is that general relativity is a field theory as any other. We should thus first try to quantize it as any other field theory. Unfortunately, it turns out that it is a non-renormalizable field theory when following the standard methods for quantum field theory. This means that we can not get any finite predictions from this method. Physically, this means that the quantum fluctuations of the gravitational field grow too fast at small scales (large energy) for the standard perturbation theory to remain valid. Thus we need an alternative quantization scheme.

\item {\bf Look for New Physical Principles}: Quantum mechanics and general relativity are two theories based on solid principles. Quantum Gravity should go beyond these theories, and thus some of these solid principles should not be true. It is actually hard to imagine which of those principles ceases to be valid and how to write a consistent physical theory if dropping one of them. This is the heart of the conceptual aspect of the research on quantum gravity: to understand the new physical principles which should underlie the theory.

\item {\bf Quanta of Space-Time}: One of those principles is to be in a continuous space-time. This basic assumption seems to be already challenged by the emergence of the Planck length $l_P$ as we have discussed above. Going further, we realize that quantizing general relativity means that the metric should be quantized and come by small discrete quanta. It is pretty hard to imagine how to reformulate general relativity in such discrete terms while still satisfying the diffeomorphism symmetry of the theory. A step in this direction has been the development of non-commutative geometries that allow for example to have a fundamental length scale while retaining the Poincar\'e symmetry. However, writing a consistent theory of a fluctuating quantized geometry will necessarily require developing further new mathematics.

\item {\bf Lack of Experimental Data}: Clearly, there are very few experimental data on quantum gravity to guide us. We can not rely on modeling experiments in order to build the theory step by step. The few possible experimental signatures of quantum gravity rely up to now on cosmological models and extreme astrophysical events. Therefore, we have to develop quantum gravity without the immediate empirical guide and to imagine realistic experimental tests, using our current technology, which could probe some aspects of quantum gravity.



\end{itemize}
These are the main big obstacles when seeking for quantum gravity. When looking more closely to one or the other approaches to quantum gravity, one will of course identify several more concrete  problems to solve on the way. But at the end of the day, to my own point of view, these are the main issues that have to be addressed.

\medskip

There have been various approaches to quantum gravity since the mid-twentieth century. There have been two main tendencies. Either we follow the ``conservative" approach whose main motivation is the straightforward quantization of general relativity. The logic is to go as far as possible using the standard methods, and see what are the smallest possible adjustments to do in order to make them work consistently. Or we try a ``top-down" approach where we postulate what we think is a good model for quantum gravity, satisfying a certain number of principles that we judge fundamental. Then we work from there, study the dynamics of the model and try to derive a semi-classical regime where we recover both general relativity and quantum mechanics. In the end, the usual way to proceed is to follow the conservative approach until we have a more or less clear idea of what the theory could be at the end of the line, and then make the jump to postulating a model of quantum gravity, from which we hope to recover general relativity as a classical limit.

Loop quantum gravity is such an approach to quantum gravity. It is based on a canonical quantization of general relativity reformulated as a gauge theory. It start with a careful analysis of the classical phase space of general relativity. It then proposes a quantization scheme defining quantum states of 3d geometry, called spin network states, and how they evolve. The goal is then to check how to recover general relativity from the final quantum theory: we need to define suitable semi-classical states and derive their effective dynamics. The strong point of this approach is the rigorous derivation of a discrete spectra for geometrical observables, such as the areas and volumes. This is a concrete implementation of the idea of quanta of geometry. The weak point of the construction is the definition of the dynamics of the quantum states. There are various ambiguities in defining the Hamiltonian at the quantum level, and there is no easy way to test and classify the different possibilities.

The spinfoam framework was introduced as an attempt to remedy to this issue and define the dynamics of loop quantum gravity in an efficient and useful way. It is a path integral formalism, inspired from the discretization of general relativity \'a la Regge and from topological quantum field theories. Spinfoams define transition amplitudes between the quantum states of geometry of loop quantum gravity, and can be interpreted as discrete quantized space-times. The goal is then to study the correlations between geometrical observables given by this path integral and to analyze their properties under the renormalization flow of spinfoam models (obtained through coarse-graining of the discretized space-time geometries).

My research project has focused on this approach to quantum gravity, especially studying the foundations and definition of the spinfoam framework and trying to extract information about the semi-classical regime of the theory. This meant both analyzing in details the quantization scheme used to derive and define spinfoam models and developing the relevant mathematical tools to study the properties of the resulting quantum amplitudes.

Of course, there exist many other approaches to quantum gravity, such as string theory, dynamical triangulations, non-commutative geometries, causal sets,\dots I have chosen to investigate the loop quantum gravity and spinfoam scheme as the mixture of ``conservative" and ``top-down" tendencies that fits with my own point of view on the research for quantum gravity.

\medskip

The line of research that I would present here is, as I explained above, the spinfoam framework for quantum gravity. I have been working on this field with various collaborators over the last few years, and more particularly with two PhD students (Valentin Bonzom and Ma\"it\'e Dupuis) during the last three years. The goal of this research program is to understand clearly how to build and define rigorously spinfoam models, to clarify their geometric interpretation and how they define a consistent proposal for a theory of quantum gravity, to investigate their (semi-)classical regime and how to extract effective information from the spinfoam amplitudes. The next step, which I am currently and actively investigating, is the coarse-graining and renormalization flow of the spinfoam path integral and its relation with the standard perturbative expansion of quantum general relativity as a quantum field theory around the flat metric.

The present manuscript is organized as follows. In a first part, I start by reviewing the spinfoam framework, the main definitions and tools, from the initial loop quantum gravity to the more recent group field theory formulation. In particular, I carefully define the most recent EPRL-FK spinfoam models. I then present two proposals to extract information about the semi-classical behavior of spinfoam models, through the reconstruction of the graviton propagator from geometric correlations and through the relation between spinfoam amplitudes and Feynman diagrams for non-commutative quantum field theories with a deformed Poincar\'e symmetry. In the second part, I present some of my original contributions to this field of research on the topics mentioned above, through a series of published papers.

\chapter{The Spinfoam Framework}

The spinfoam framework proposes a discretized and regularized path integral for quantum gravity. It was originally introduced as a ``sum-over-surfaces" formalism to implement the dynamics of loop quantum gravity and define transition amplitudes for its spin network states of quantum geometry \cite{rr1}. It was then realized that there is a strong relation between this ``sum-over-surfaces" path integral formalism and state-sum models obtained particularly from discretizing the path integral for topological field theories of the BF type. This led to the creation of {\it spinfoam models} \cite{sfbaez}. This link between gravity and topological BF theory at the classical level, and between spinfoam amplitudes and state-sum models at the quantum level, was realized particularly early in three space-time dimensions \cite{sfpr_loop}: the Ponzano-Regge model for 3d quantum gravity, defined in 1968 \cite{pr}, is actually the first historical example of spinfoam models. Following this line of thoughts, some spinfoam models for quantum gravity in four space-time dimensions were proposed, initially derived using geometric quantization tools: these are the Barrett-Crane models for Euclidean and Lorentzian 4d gravity \cite{bc1,bc2}. Parallelly, the reformulation of general relativity as an ``almost-topological" field theory, expressed as a constrained BF theory, was actively investigated (see e.g. \cite{fdp}). This helps achieving a better understanding of the foundations of spinfoam models and their geometric interpretation, which leads recently to the introduction of the new EPRL-FK class of spinfoam models \cite{ls1,epr,fk,ls2,eprl}. These spinfoam models, as well as the Barrett-Crane models, can be actually derived as exact path integrals for discretized actions for constrained BF theories \cite{actionfk, actionfc,actionlb}. Finally, spinfoam models can be considering as merging loop quantum gravity, topological quantum field theory and Regge calculus for discretized general relativity.

At the same time as developing new spinfoam models, a reformulation of spinfoam models as generalized matrix models was introduced. These {\it group field theory} \cite{gft} techniques were first developed for the Barrett-Crane models \cite{fdpkr}. Then it was realized that they provide a non-perturbative definition for all (local) spinfoam models \cite{gftrr}, such that the partition function of the group field theory generates the path integral over all topologies and geometries similarly to the way that matrix models generate sums over all geometries of 2d (triangulated) surfaces.

I will review this whole spinfoam framework in the following chapter. I will describe all the details of the formalism and review all the literature of this field but I will focus on the main features of the framework.

\section{The Origin: Loop Gravity and Spin Network States}

Loop quantum gravity (LQG) is a canonical framework for a non-perturbative and background-independent quantization of general relativity. By ``canonical", it is meant that we start with a space+time splitting then we identify the dynamical variables that will evolve in time. The classical framework is the defined by the phase space formed by these dynamical variables. The quantization program consists in defining proper wave-functions over this classical phase space. These wave-functions define quantum states of geometry for general relativity. The final step is to describe how these states of quantum geometry evolve in time.

By ``non-perturbative", we mean that the formalism is not based on the study of small (graviton-like) variations of the metric around the flat space-time metric, as is the standard quantum field theory perturbative approach to general relativity. Instead, loop quantum gravity will define basic excitations of the gravitational field arbitrarily far from the flat metric and will describe the quantum space-time directly at the Planck scale, where the geometry comes by ``quanta".  By ``background independence", we implicitly mean two different but related things. First, LQG does not rely on a particular background metric, around which we perturb. Actually, we can view LQG as an expansion of the gravitational degrees of freedom from the no-metric state, where the metric is completely degenerate and vanishes (thus defining the space-time as a totally topological concept). Thus one main issue in LQG is to re-construct non-trivial classical metric such as the flat Minkowski metric or cosmological metrics. Indeed, defining a stable phase of LQG with the gravitational field fluctuating around the flat metric is a completely non-perturbative effect in this framework. Second, ``background independence" also refers to the diffeomorphism invariance of LQG. Indeed, the invariance of general relativity under space-time diffeomorphisms is considered as fundamental and strongly encoded and implemented in the LQG framework. This is the logical consequences of considering symmetries as fundamental for quantum field theories (e.g. gauge field theories defining our standard model for particle physics) and simply physics in general. The resulting problem for LQG is the loss of the simple concept of localized space-time points. Localization on a state of quantum geometry has to be reconstructed a posteriori based on purely relational terms (we need to localize something with respect to other systems) and it does not help that we do not have a (non-degenerate) background metric to help us define a space-time coordinate systems.

In this section, I will review the loop quantum gravity framework and its spin network states for quantum geometry, avoiding as many technicalities as possible. The interested reader can refer to \cite{LQGc,LQGt} for more detailed and thorough reviews of the formalism.

\subsection{A Canonical Framework for Gravity as a Gauge Theory}

Loop quantum gravity is based on the first order formulation of general relativity. Starting with the space-time manifold $\cM$, instead of describing its geometry in term of the metric $g_\mn(x)$, we define it in term of the tetrad $e^I_\mu(x)$ and Lorentz connection $\om^{IJ}_\mu(x)$, where $\mu,\nu$ are space-time indices referring to a coordinate system $x_\mu$ on $\cM$ while $I,J$ are (internal) Minkowski indices referring to coordinates on the tangent space to $\cM$. Let us recall that the metric $g_\mn$ defines the infinitesimal distance element:
\be
ds^2\,=\, g_\mn\,dx^\mu dx^\nu,
\ee
which allows to define distances and geodesics on $\cM$. On the other hand, the tetrad $e^I_\mu$ is a 1-form on $\cM$. It gives the local map between the coordinate system on $\cM$ and the coordinate system on its tangent space. It relates the metric $g_\mn$ to the flat metric $\eta_{IJ}$ on the Minkowski space-time, and thus diagonalizes $g_\mn$ (remember that any real symmetric matrix is diagonalizable):
\be
g_\mn\,=\,
e^I_\mu e^J_\nu\,\eta_{IJ}.
\ee
In more geometric terms, the tetrad defines the local orthonormal basis/frame on the space-time manifold. Then the connection $\om^{IJ}_\mu(x)$ defines the parallel transport of this frame along the manifold.

In those variables, the Palatini-Holst action for general relativity is \cite{holst}:
\be
S[e,\om]=-\f1{l_P^2}\int_{\cM}
\f12\,\epsilon_{IJKL} e^I \w e^J \w F^{KL}[\om]
+\f1\gamma\, e^I \w e^J \w F_{IJ}[\om].
\label{holst}
\ee
The 2-form $F[\om]\,\equiv\, d\om + \om\w\om$ is the curvature tensor of the connection $\om$ and $\w$ is the wedge product for differential forms on $\cM$. The coupling constant $\gamma$ is the Immirzi parameter. It is in front of the Holst term \cite{holst}. It does not affect the classical equations of motion of gravity -even though it affects the coupling of fermionic fields to gravity \cite{immfermion}- but becomes relevant at the quantum level.

Let us start by ignoring the Holst term. Then the equation of motion for the connection reads $d_\om e\w e\,=0$, which is equivalent to $d_\om e=0$ as long as the tetrad field is non-degenerate $\det e\ne 0$. This equation of compatibility between the tetrad $e$ and the connection $\om$ implies that $\om$ is the Levi-Civita connection defined as a function of $e$. In mathematical term, it means that the torsion vanishes. Then the equation of motion for the tetrad reads $\epsilon_{IJKL} e^J \w F^{KL}\,=0$, which reduces to the Einstein equations imposing the vanishing of the Ricci curvature tensor $R_\mn=0$ as soon as we require a vanishing torsion $d_\om e=0$.
Now coming back to full action including the Holst term, the equation imposing the vanishing of the torsion is not affected by this new term as long as the tetrad is non-degenerate. Then it contributes to the second equation of motion a term $e^J \w F_{IJ}$, which actually vanishes when the torsion is zero. Thus, assuming that the tetrad (and therefore the metric) is non-degenerate, we have a full equivalence at the classical level between this Palatini-Holst action and the standard formulation of general relativity defined in term of the metric and the Einstein equations. Moreover, the Holst term with the Immirzi parameter $\gamma$ is not relevant as this level. Let us nevertheless point out that the physics/mathematics of the degenerate sector, and the relevance of the Holst term in this regime, is not yet fully understood.

Finally, this Holst-Palatini action is invariant under both space-time diffeomorphisms (acting on space-time indices) and local Lorentz transformation (acting on internal indices). Let us also point out that the cosmological constant $\Lambda$ would be simply taken into account by inserting a volume term to this action $+\Lambda\,\eps_{IJKL} e^I\w e^J \w e^K\w e^L$.

\medskip

As we said earlier, loop gravity is based on a canonical analysis of general relativity in its first order formalism. We thus choose a space-time manifold with a simple topology $\cM =\R\times \Sigma$ where $\R$ is the time dimension and $\Sigma$ is the canonical hypersurface defining space. Then we identify the dynamical variables, which enter the action with time derivatives. Finally, it is possible to write the full action in the following manner (e.g. \cite{holst,barros}):
\be
S[E,A,\lambda,N]\,=\,
\f1{l_P^2}\int dt\int_\Sigma d^3x\,
\f1\gamma A^i_a\pp_0 E^a_i - H,
\label{actionLQG}
\ee
where the kinetic term defines the Poisson bracket and the Hamiltonian consists only in constraints:
\be
H=\lambda^i G_i + N^a V_a + N S,
\ee
with three different sets of constraints
\bes
&&G_i\,=\,
D_aE^a_i= \pp_a E^a_i+\eps_{ij}{}^kA^j_aE^a_k, \\
&&V_a\,=\,E^b_i F^i_{ab}, \\
&&S\,\,\,=\,
\f{E^a_iE^b_j}{\sqrt{\det E}}\,\left(
\eps_k^{ij}F^k_{ab}+2(1-\gamma^2)K^i_{[a}K^j_{b]}.
\right)
\ees
Here $i,j,k$ run from 1 to 3 and are internal spatial indices. Similarly, $a,b,c$ are space indices referring to a coordinate system on the hypersurface $\Sigma$.
There are a lot of new notations in this new action, so let's introduce them slowly.

First, the fundamental canonical variables are the triad $E$ and the Ashtekar-Barbero connection $A$, which satisfy the kinematical Poisson bracket:
\be
\{A^j_b(x),E^a_i(y)\}\,=\,
\gamma\,\delta^j_i\delta^a_b\delta^{(3)}(x-y),
\ee
which depends explicitly on the Immirzi parameter $\gamma$.
The triad $E^a_i$ is constructed from the spatial components of the tetrad field:
\be
E^a_i=
\f12 \eps^{abc}\eps_{ijk} e^j_b e^k_c.
\ee
Furthermore, in order to derive the action above \Ref{actionLQG}, we have assumed the {\it time gauge}, $e^0_a=0$, geometrically meaning that the time normal to the canonical hypersurface $\Sigma$ is always taken fixed to $n^I=(1,0,0,0)$. This time gauge condition actually breaks the original local Lorentz invariance of the action down to a $\SU(2)$ invariance under local 3d rotations of the spatial frame. Then the conjugate variable to the triad is the Ashtekar-Barbero connection $A^i_a$, which is constructed in term of the spin-connection $\Gamma(E)$ and the extrinsic curvature $K$:
\be
A^i_a=\Gamma^i_a(E)+\gamma K^i_a,
\ee
with $K^i_a=\om^{0i}_a$ and the spin-connection $\Gamma(E)$ defined in term of the triad $E$ through the equation:
\be
\eps^{abc}\,\left(
\pp_a e^i_b + \eps^i{}_{jk}\Gamma^j_a e^k_b
\right)=0.
\ee
The triad $E^a_i$ and the extrinsic curvature $K_a^i$ are $\su(2)$-valued 1-forms on $\Sigma$. Since $\Gamma(E)$ is a function of only $E$ (and does not depend on $A$), we see that $E$ and $K$ actually also provide us with a set of conjugate variables:
\be
\{K^j_b(x),E^a_i(y)\}\,=\,
\delta^j_i\delta^a_b\delta^{(3)}(x-y),
\ee
which does not involve the Immirzi parameter $\gamma$. Nevertheless, it is the extra-term $\Gamma(E)$ in the definition $A$ that makes that the field $A^i_a$ is actually a $\su(2)$-valued connection on $\Sigma$, thus leading to a canonical formulation of general relativity as a ($\SU(2)$) {\it gauge field theory}. The change of variable $K^i_a\arr A^i_a$ is actually a canonical transformation at the classical level, which turns out to not be implemented by a unitary transformation at the quantum level \cite{immRT}.

\medskip

Then gravity is a totally constrained system, which means that the Hamiltonian $H$ is made of only constraints and vanishes on-shell (i.e on solutions to the classical equations of motion). These constraints $G_i,V_a,S$ form a first-class constraint system and thus generate symmetries of the action.  First, the Lagrange multipliers $\lambda_i$ impose what we call the Gauss law constraints
$G_i=D E_i$. Their Poisson brackets are very simple:
\be
\{G_i,G_j\}\,=\,-\eps_{ij}{}^kG_k\,,
\ee
and form a $\su(2)$ Lie algebra.
These $G_i$'s actually generate $\SU(2)$ gauge transformations on the variables $E$ and $A$, which read as:
\be
\left|
\begin{array}{lcl}
E^a &\arr & g E^a g^{-1}\\
A_a &\arr & g A_a g^{-1} +g\pp_a g^{-1}
\end{array}
\right.\,,
\ee
where $g$ is a $\SU(2)$ field on $\Sigma$ parameterizing the $\SU(2)$ transformations.
These transformations act on the tangent space to $\Sigma$ and thus act on the internal index $i$.
These Gauss law constraints appear because we are working in a first order formulation in term of vierbein/connection variables. If we were working in the second order formulation in term of the metric field, we would not have internal indices and the ADM canonical analysis would only lead to the constraints $V_a$ and $S$. Nevertheless, it is these $G_i$ constraints imposing a $\SU(2)$ gauge invariance which show clearly that loop gravity is based on a gauge theory reformulation of general relativity.

The next constraints are respectively the vector constraints $V_a$ and the scalar constraint $S$. Together they generate space-time diffeomorphisms. The Lagrange multiplier are respectively the shift $N^a$ and the lapse $N$. I will not give details on the Poisson brackets of these constraints. The motivated reader will find more details in e.g. \cite{LQGt}. There are a couple of subtleties I will nevertheless mention. First, to truly get the generators of the diffeomorphisms and a closed Poisson-Lie algebra, we need to introduce terms proportional to the Gauss law constraints to $V_a$ and $S$. Second, $V_a$ and $S$ only truly generate space-time diffeomorphisms on fields solution to the classical equations of motion (see e.g. \cite{LQGt,thomasmeasure}). These two subtleties lead to issues during the quantization process, but we will not discuss them here.

\medskip

This concludes the description of the classical setting of loop gravity. I will describe below the quantization programme of loop quantum gravity. The goal is to define suitable wave-functions of the geometry and implement the constraints imposing the invariance under $\SU(2)$ gauge transformations and space-time diffeomorphisms on the resulting Hilbert space of quantum geometry states.


\subsection{The Hilbert Space of Spin Network States}

Once the classical setting is fixed and the phase defined as explained above, the quantization programme of loop quantum gravity is rather straightforward. It is performed in a few key steps, as follows.

\begin{enumerate}

\item Our canonical variables, describing the geometry on the 3d spatial slice $\Sigma$, are the triad $E$ and the Ashtekar-Barbero $\SU(2)$-connection $A$. Our choice of polarization is simple: we choose wave-functions $\psi(A)$ which depend on the connection. Then the triad $E$ will be represented as a differential operator on such wave-functions.

\item We refine this ansatz by choosing a particular set of observables for the theory. We indeed focus on the algebra of holonomy-flux observables. These are natural observables for a gauge field theory, they allow to parameterize the whole space space and to write easily gauge-invariant observables (like Wilson loops). More precisely, we define the holonomies by integrating the connection $A$ along links $e$ embedded in $\Sigma$ and we define flux observables by integrating the triad $E$ against test functions on surfaces $\cS$:
    \be
    U_e[A]\,\equiv\,\cP\,e^{\int_e ds\,A^i_a J_i\pp_sc^a},\qquad
    E_\cS[f]\,\equiv\,\int_\cS\, E_i f^i,
    \ee
    where $J_i$ are the generators of the $\su(2)$ Lie algebra, $c:s\in[0,1]\mapsto c^a(s)\in e\subset\Sigma$ is the map defining the curve $e$ and $f:\cS\arr\su(2)$ is a $\su(2)$-valued test function on the surface $\cS$.

    Now, we choose {\it cylindrical wave-functions} which only depend on the connection $A$ through a finite number of degrees of freedom. To do so, we choose a graph $\Gamma$ embedded in the canonical hypersurface $\Sigma$ and we consider functions of the holonomies $U_e[A]$ along its edges $e\in\Gamma$:
    \be
    \psi(A)\,\equiv\,\psi_\Gamma(\{U_e[A]\}_{e\in\Gamma}).
    \ee
    We choose wave-functions which are {\it invariant under $\SU(2)$ gauge transformations}, so that they solve the Gauss law constraints $G_i$.
    It is possible to define a measure and a scalar product on this space of wave-functions. This defines the Hilbert space $\cH_\Gamma$ of $L^2$ cylindrical functions based on the graph $\Gamma$. A basis on this space $\cH_\Gamma$ is provided by the {\it spin network states} as we will explain later.

    We then define our kinematical Hilbert space as the sum of these spaces $\cH_\Gamma$ over all possible graph $\Gamma$:
    \be
    \cH_{kin}\,\equiv\,``\bigoplus_\Gamma" \cH_\Gamma.
    \ee
    This is actually a subtle point due to two reasons. First, we must carefully choose the mathematical properties of graphs over which we sum. Second, this is not strictly a direct sum over the graphs $\Gamma$, but more precisely a {\it projective limit}. Indeed, we consider all wave-functions based on all possible graphs but we identify cylindrical functions on a graph $\Gamma$ as functions on on all possible larger graphs $\Gamma'$ which contain $\Gamma$ as a sub-graph. Then it is possible to define a measure on this space of function, called the Ashtekar-Lewandowski measure, and to define the space $\cH_{kin}$ as $L^2$-functions with respect to that measure \cite{projective}.

    Finally, it was shown that this representation of the  algebra of holonomy-flux observables $U_e[A],E_\cS[f]$ on the Hilbert space $\cH_{kin}$ of spin networks, with holonomies acting by multiplication and flux acting as derivative operators, is unique if requiring a proper representation of space diffeomorphisms acting on this space $\cH_{kin}$. This is the LOST theorem, one of the founding theorems of loop quantum gravity \cite{lost}.

\item The next step is to solve the vector constraints $V_a$, which generate space diffeomorphisms on the canonical hypersurface $\Sigma$. The action of spatial diffeomorphisms on the wave-functions $\psi_\Gamma(\{U_e\}_{e\in\Gamma})$ is the natural one: it simply shifts the embedded graph $\Gamma$ in $\Sigma$. Since the Hilbert spaces $\cH_\Gamma$ are isomorphic to each other when the graphs $\Gamma$ defer by a diffeomorphism, it is straightforward to implement the invariance under spatial diffeomorphisms by considering equivalence classes of graphs under diffeomorphisms:
    \be
    \cH_{diff}\,\equiv\,``\bigoplus_{[\Gamma]}" \cH_\Gamma,
    \ee
    where the ``" means as before that we identify the states based on a graph $\Gamma$ as states living on all larger graphs containing $\Gamma$. The diffeomorphism-invariant states in $\cH_{diff}$ obviously solve the vector constraints $V_a$ at the quantum level.

    Here too, there are a couple of subtleties and ambiguities in the procedure. First, let us start by pointing out that the kinematical Hilbert space $\cH_{kin}$ is not separable. The hope is that quotienting by the spatial diffeomorphisms allows to obtain a {\it separable Hilbert space $\cH_{diff}$} with a countable basis of quantum states. This actually depends on the specific choice of graphs $\Gamma$ and the exact class of diffeomorphisms whose action we consider. The standard choice is to consider piecewise linear graphs, but there exist many possible variations of the class of graphs. The problem is then that the equivalence classes of such graphs under diffeomorphisms contain a continuous moduli when the graph contains nodes with a valence higher or equal to 5. A solution to cure this issue is to consider maps which are differentiable everywhere but at a finite number of points as proposed in \cite{diffwinston}, then we obtain as expected a separable Hilbert space $\cH_{diff}$ with the equivalence classes $[\Gamma]$ determined only by the combinatorial structure of the graph $\Gamma$ and the topological winding numbers of $\Gamma$ embedded in $\Sigma$.

    The second more important subtlety is whether to truly consider embedded graph or to move to working with non-embedded graph. From the strict point of view of the loop gravity quantization, we must work with embedded graph and equivalence classes labeled by both the combinatorial structure of the graph and the topological information defining the relation of the graph with the underlying space-time manifold. However, from the larger perspective of a ``top-down" approach, we would like to argue that it seems more appropriate to consider non-embedded graph. Indeed, a totally background-independent quantum gravity theory would like to get rid of every a priori geometrical and topological information about the space-time manifold. The topology and geometry of the quantum space-time should be fully reconstructible from the quantum state. More precisely, a spin network based on a combinatorial graph (without further topological data) seems to contain enough information to define the relations between space points and thus define the topology of the spatial hypersurface. This is actually the point of view developed in {\it spinfoam models}: the spin network state based on a non-embedded graph define the 3d boundary state, then inserted in the path integral, it determines (in the semi-classical regime) the 4d bulk structure, which finally determines the space-time topology and induces the space topology.

\item The final step is to solve the scalar constraint $S$, also called the Hamiltonian constraint since it generates time re-parameterizations. This is the final goal of the LQG program: derive the physical Hilbert space $\cH_{phys}$ of quantum states that solve all constraints $G_i,V_a,S$. There are a few proposal for a regularized Hamiltonian constraint at the quantum level, with the main one being Thiemann's proposal \cite{thomas_dyn}. The problem is that we do not yet know how to test such a proposal to check whether it defines the correct dynamics for the theory: once a Hamiltonian constraint operator proposed, we do not know how to solve it and identify its null subspace, we do not know to reconstruct coherent semi-classical wave-packets peaked on some classical geometry and we do not know how to coarse-grain the spin network states to extract their large scale structure. This is where the LQG program gets stuck. But as we will discuss later, in section \ref{stuck}, there are alternatives to the original LQG which allow to address the issue of constructing the dynamics for the theory.

\end{enumerate}

\medskip

Now that we have reviewed the main of the loop quantum gravity programme, we need to introduce the main tool of the LQG formalism: its {\it spin network states}.

Let us start with an oriented graph $\Gamma$, with $E$ edges and $V$ vertices. For each edge $e$, we will call its source vertex $s(e)$ and its target vertex $t(e)$. The cylindrical functions based on $\Gamma$ depends on the holonomies $U_e[A]\in\SU(2)$ along the graph's edges:
$$
\psi(A)\equiv\psi_\Gamma(\{U_e[A]\}_{e\in\Gamma}).
$$
We would like to further  require that these functionals be invariant under $\SU(2)$ gauge transformations. These transformations act very simply on the holonomies, they act only at the end vertices of the considered edge:
$$
U_e\,\arr\,h_{s(e)}U_e h_{t(e)}^{-1},\qquad h_{s(e)},h_{t(e)}\in\SU(2).
$$
Thus we consider gauge invariant cylindrical functions:
\be
\psi_\Gamma(\{U_e\})\,=\,\psi_\Gamma(\{h_{s(e)}U_e h_{t(e)}^{-1}\}),\qquad
\forall\,\,h_v\in\SU(2)^{\times V},
\ee
which are then functions on the coset space $\SU(2)^{\times E}/\SU(2)^{\times V}$. Indeed, they depend on $E$ $\SU(2)$ group elements living on the graph's edge and are invariant under the action of $\SU(2)$ at the graph's $V$ vertices.

We endow this space of function with the invariant Haar measure on $\SU(2)$, thus defining a scalar product:
\be
\la \psi_\Gamma|\tpsi_\Gamma\ra\,=\,
\int_{\SU(2)^{\times E}} \prod_e dU_e\,
\overline{\psi}(\{U_e\})\,\tpsi(\{U_e\}).
\ee
And we can define our Hilbert space $\cH_\Gamma$ of gauge invariant cylindrical wave-functions based on the graph $\Gamma$ as the $L^2$-space with respect to this measure:
\be
\cH_\Gamma=\,L^2\left(\SU(2)^{\times E}/\SU(2)^{\times V}\right).
\ee
This measure, and thus scalar product, only depend on the combinatorial structure of the graph $\Gamma$. In particular, the spaces $\cH_\Gamma$ are isomorphic for two graphs in the same equivalence class under spatial diffeomorphisms.

In order to obtain a basis of quantum states on $\cH_\Gamma$, we use the Peter-Weyl theorem which states that every $L^2$-function over $\SU(2)$ can be decomposed on the irreducible representations of $\SU(2)$. This harmonic decomposition reads:
\be
\vphi(g)\,=\,\sum_{j\in\N/2} \tr \,\vphi^j\,D^j(g).
\ee
The half-integer $j$ is the {\it spin} and labels the irreducible representations of $\SU(2)$. We call $V^j$ the Hilbert space attached to this representation. Its dimension is $d_j=\dim V^j=(2j+1)$. The matrix $\vphi^j$ is $d_j\times d_j$ and contains the ``Fourier" components of $\vphi$. The Wigner matrix $D^j(g)$ represents the group element $g$ in the $j$representation of spin $j$. We can also write it in the standard bra-ket notation:
\be
D^j_{ab}(g)\,=\,\la j,a|g|j,b\ra\,,
\ee
where $a,b$ labels a basis of $V^j$ (usually the magnetic moment basis diagonalizing the $J^z$ operator). Finally the trace is taken over $V^j$. The reverse formula is easily obtained by using the orthogonality of the Wigner matrices with respect to the Haar measure:
\be
\int_{\SU(2)} dg\, D^j_{ab}(g)D^k_{cd}(g^{-1})\,=\,
\int_{\SU(2)} dg\, D^j_{ab}(g)\,\overline{D^k_{dc}(g)}\,=\,
\f{\delta_{jk}}{d_j}\,\delta_{ad}\delta_{bc}.
\ee

Using this Peter-Weyl decomposition, we easily show that a basis of $\cH_\Gamma$ is provided by the so-called spin network states. They are labeled by one spin $j_e\in\N/2$ on each edge and one intertwiner state $i_v$ at each vertex. An intertwiner $i_v$ is, as its name suggests, a $\SU(2)$-invariant map which intertwines all the representations on the edge attached to the vertex $v$. More precisely, it is a map:
\be
i_v:\bigotimes_{e|s(e)=v} V^{j_e}\longrightarrow \bigotimes_{e|t(e)=v} V^{j_e},
\ee
which commutes with the $\SU(2)$-action, i.e $i_v\,g\,|\phi\ra=g\,i_v\,|\phi\ra$ for all states $\phi$ and all group elements $g\in\SU(2)$. Then the spin network functional is defined by contracting the intertwtiners $i_v$ along the edges $e$ using the holonomies living on these edges:
\bes
\psi^{j_e,i_v}_\Gamma(\{g_e\})
&\equiv&
\tr\,\bigotimes_v i_v\,\bigotimes_e D^{j_e}(g_e)\\
&=&
\sum_{m^s_e,m^t_e}\prod_e \la j_e m^s_e |g_e | j_e m^t_e \ra\,\prod_v \bigotimes_{e|t(e)=v}\la j_e m^t_e |\, i_v \,\bigotimes_{e|s(e)=v}| j_e m^s_e\ra,\nn
\ees
where the $\tr$ is take over all representation spaces $V^{j_e}$ living on the graph's edges.
Now to truly obtain a basis of $\cH_\Gamma$, we simply need to choose a basis of intertwiner states at each vertex.

To understand the meaning of the intertwiners, we can put all representations on the same side. Then an intertwiner is a $\SU(2)$-invariant linear map:
\be
i_v:\bigotimes_{e|s(e)=v} V^{j_e}\otimes\bigotimes_{e|t(e)=v} \overline{V}^{j_e}
\longrightarrow \C.
\ee
This is simply a {\it singlet} state. Moreover, in the special case of a 3-valent vertex, once the three representations around the vertex are fixed, there exists a unique intertwiner state (up to a global factor), which is simply given by the Clebsh-Gordan coefficients. Going further, considering a vertex with arbitrary valency, we can always decompose into basic 3-valent blocks and express all possible intertwiner states in term of the basic Clebsh-Gordan coefficients.

These spin network states are essential to loop quantum gravity, especially because they diagonalize the geometric operators such as areas and volumes.

\subsection{Geometric Observables and Discrete Spectra for Area/Volumes}

Back to the classical level, geometric observables on the hypersurface $\Sigma$ are functions of the 3-metric ${}^3h_{ab}$ on this spatial slice. This 3-metric can be entirely expressed in term of the triad $E^a_i$, which allow to write all geometric observables as functions of $E$. We focus in particular on areas and volumes, which are obviously $\SU(2)$ invariant observables (but not diffeomorphism-invariant). For a surface $\cS\subset\Sigma$, with coordinates $\sigma^1,\sigma^2$, the area can be written as (see e.g.\cite{primer}):
\be
\cA_\cS=
\int d\sigma^1d\sigma^2\,\sqrt{
\eps_{abc}\f{\pp x^a}{\pp\sigma^1}\f{\pp x^b}{\pp\sigma^2}E^c_i(x(\sigma))
\eps_{def}\f{\pp x^d}{\pp\sigma^1}\f{\pp x^e}{\pp\sigma^2}E^f_i(x(\sigma))
}\,.
\ee
As for the volume of a region $\cR\subset\Sigma$, we get:
\be
\cV_\cR=
\int dx^3\sqrt{\f1{3!}\eps_{abc}\eps^{ijk}E^a_i(x)E^b_j(x)E^c_k(x)}\,.
\ee
These are turned into operators acting on quantum states by quantizing the triad $E^a_i$ as the derivative operator with respect to the connection $A_a^i$ as expected from the Poisson bracket:
\be
\wh{E}^a_i(x)\,\equiv\,
il_P^2\,\gamma \,\f{\pp}{\pp A_a^i(x)},
\ee
where $\gamma$ is the Immirzi parameter.

I will not describe here the details of the derivation of the quantum operators $\wh{\cA}$ and $\wh{\cV}$. The interested reader can refer to \cite{LQGc,LQGt}. The quantization relies on three basic steps: split the integrals as Riemann integrals into elementary pieces on which the fields can be considered as (almost) constant, quantize $E$ as above, regularize the product of $E$-fields by point-splitting. Beyond the quantization process, what is interesting is the resulting geometrical picture.

Considering a particular spin network state $\psi_\Gamma$ in $\cH_{kin}$ based on the graph $\Gamma\subset\Sigma$, the area operator $\wh{\cA}_\cS$ gets contributions for all the edges $e$ that puncture the surface $\cS$ while the volume operator $\wh{\cV}_\cR$ gets contributions only from the vertices $v$ (with a valency higher or equal to 4) contained in the region $\cR$.

Starting with the area, the area operators $\wh{\cA}_\cS$ are actually diagonalized  by the spin network states. More precisely, looking at the action of the differential operator $\f{\pp}{\pp A_a^i(x)}$ on a holonomy $U_e[A]$, we see that the operator $\wh{E}_i(x)$ for $x\in e\in\Gamma$ acts on a spin network state by insertion of the $\su(2)$ generator $J_i$ along the edge $e$ at the point $x$. Thus, looking at an elementary surface $\cS$ which only intersects the graph $\Gamma$ at a single point located on the edge $e$, we can easily compute the action of the corresponding area operator:
\be
\wh{\cA}_\cS\,\psi^{j_e,i_v}_\Gamma\,=\,
\gamma \,l_P^2\,\sqrt{J^e_iJ^e_i}\,\psi^{j_e,i_v}_\Gamma
\,=\,
\gamma \,l_P^2\,\sqrt{j_e(j_e+1)}\,\psi^{j_e,i_v}_\Gamma,
\qquad
\forall \cS \,\textrm{s.t.} \,\cS\cap\Gamma =\{x\in e\},
\ee
where the index $e$ in $J^e_i$ refers to the $\SU(2)$ action on the edge $e$.
Then, considering a generic surface, we decompose it into such elementary patches, which intersect the graph at a single point, and we define its total area as the sum of the areas of those elementary patches. This way, we have derived the whole area spectrum for loop quantum gravity
\be
\{\gamma \,l_P^2\, \sum_{i=1}^N \sqrt{j_i(j_i+1)},\quad\forall N,\forall j_i\in\N/2\}.
\ee
The first remark is that we have obtained a {\it discrete spectrum} for the area of surfaces, thus confirming the initial idea that geometry comes in discrete quanta in quantum gravity. The second remark is that the Immirzi parameter, which was not relevant at the classical level, seems to play a crucial role in the quantum theory, since it comes at a numerical factor in front of the area eigenvalues. Finally, a last remark before moving on to the volume operator is that there are actually quantization ambiguities in defining the square-root operator $\sqrt{J^e_iJ^e_i}$. More precisely, these are simple usual ordering ambiguities. It turns out that, although $\sqrt{j_e(j_e+1)}$ is the standard area spectrum in loop quantum gravity, we could also derive the equidistant spectrum, simply $j_e$, with a slightly different ordering e.g. \cite{krasnov_area}.
These various area operators differ in their orderings. They should be relevant to different physical/mathematical situations or one could argue that one should identify a definitive normal ordering. These quantization ambiguities are still an open issue in loop quantum gravity.
%

\medskip

Now, considering the volume operator $\wh{\cV}_\cR$, it involves triple products of operators $E^a_i(x)E^b_j(x)E^c_k(x)$ acting at the same point and involving three different directions (due to the tensor $\eps_{abc}$). Due to this requirement of having three different directions, a point $x$ needs to be a (non-bivalent) vertex $v$ of the graph. As before for the area, let us focus on an elementary chunk of 3d space which contains a single vertex $v\in\Gamma$. Then the volume operator can be considered as acting directly on the intertwiner state $i_v$ living at the considered vertex $v$ and selects triplets of edges attached to $v$:
\be
\wh{\cV}_\cR\, i_v\,=\,
\gamma^{\f32}\,l_P^3\,
\sum_{(e,f,g)} \sqrt{\f18\f{\eps^{ijk}}{48}J^e_iJ^f_jJ^g_k}\,\, i_v.
\ee
This expression vanishes for 3-valent vertices. The simplest non-trivial case is for a 4-valent vertex, in which case the sum over triplets of edges drops out and we are left with a single contribution. Then we can look for eigenvectors of this volume operator and build spin network states diagonalizing the volume operators as well as the area operators. However, despite a lot of work of the volume operator (see e.g.\cite{vol_lqg}), the full spectrum of this operator is not entirely explicitly known. At least, it is clear that we have derived a discrete spectrum once again. It is nevertheless an important question since understanding more about the volume operator is essential to both understanding the deep geometrical meaning of spin network states and to studying the dynamics of the theory since the Hamiltonian constraint operator involves directly the volume operator (see e.g. \cite{thomas_dyn}).

There are even more subtleties on the volume operator than for the area operator. First, the operator $\eps^{ijk}J^e_iJ^f_jJ^g_k$ is Hermitian but not positive. Therefore, we truly need to consider the absolute value $|\eps^{ijk}J^e_iJ^f_jJ^g_k|$ in the expression above; or else manage to identify and separate the positive and negative contributions to this operator. Second, depending on the regularization scheme, there are various proposals for an explicit volume operator. The main two possibilities are whether we sum over the triplets of edges outside the square-root or inside it. This actually makes a big difference in the algebraic and spectral properties of the volume operator. The interested reader will find a comprehensive study and extensive discussion of these possibilities in \cite{thomas_vol}.

\medskip

What's to be remembered of this construction of operators for geometric observables such as areas and volumes is that, first, loop quantum gravity provides a (mathematically) rigorous quantization for these operators and, second, the spectra of these geometric operators turns out to be discrete in Planck units. This leads to a picture of a {\it discrete quantum geometry}.

Indeed, it seems that we can propose a straightforward geometric interpretation of spin network states: the edges $e$ of the graph are dual to elementary surfaces whose area is given by the spin $j_e$ carried to the edge, and the vertices $v$ are dual to elementary chunks of 3d space bounded by those elementary surfaces and whose volume is determined by the intertwiner $i_v$ living at the vertex. More precisely, this interpretation hints towards a possible reconstruction of a discrete geometry dual to the spin network state, with (classical) polyhedra reconstructed around each vertex whose faces are dual to the edges attached to the vertex and whose exact shape would depend on the explicit intertwiner living at the vertex. This point of view has been particularly developed from the perspective of geometric quantization. Indeed, it is possible to see intertwiners  as quantum polyhedra. In particular, much work  has focused on the interpretation of 4-valent intertwiners as quantum tetrahedron \cite{qtetra,qtetrabb}. This point of view has been particularly useful to built spinfoam models as quantized 4-dimensional triangulations \cite{bc1,bc2,qtetrabb}.

Nevertheless, in order to set this interpretation on stronger foundations and truly identify intertwiners as quantum polyhedra and spin network states as discrete geometries, we need to be able to build semi-classical intertwiner states whose shape would be peaked on classical polyhedra and then to glue them together in order to build semi-classical spin network states peaked on classical discrete geometries. There has actually been a lot of research work done on developing such concepts, e.g. complexifier coherent states introduced by Thiemann \cite{coh_thomas,coh_thomas2}, the related holomorphic spin network states \cite{coh_claudio}, the coherent intertwiner states which I introduced with Speziale and collaborators \cite{ls1,coh_fkl}. Particular recent lines of research which seems to re-unify these works and viewpoints are the twisted geometry framework \cite{twistedgeom} and the $\U(N)$ framework for intertwiners \cite{un}, which actually converge themselves to a unified picture of coherent spin network states as semi-classical discrete goemetries \cite{spinorlqg}. These frameworks are partly inspired from the picture of coherent intertwiners \cite{ls1,coh_fkl} and allow to define explicit variables which control the shape of intertwiners and also parameterize classical polyhedra, thus creating an explicit bridge between the two.

\subsection{Where the LQG formalism gets stuck: the Dynamics}
\label{stuck}

So up to now, we have seen how the loop quantum gravity program quantizes the classical phase space of the triad and $\SU(2)$-connection variables using cylindrical wave-functions and spin network states. We have explained how to raise geometric observables to quantum operators and reviewed how the spin network states solve the Gauss law constraints implementing the $\SU(2)$-gauge invariance and the vector constraints imposing the invariance under spatial diffeomorphisms. The last step is to take care of the Hamiltonian constraint, which generate time re-parameterization and more generally space-time diffeomorphisms (together with the vector constraints) and which encodes the dynamics of the theory. And this is the hard step, where the LQG machinery gets stuck. This is due to two related issues.

The first issue is quite generic to general relativity and quantum gravity. It is the fact that the Hamiltonian vanishes and that we have a Hamiltonian constraints implementing the invariance of the theory under translations in time. This is commonly called the ``problem of time": in order to recover a real dynamics with an actual evolution in time, one has to find a suitable internal time variable (clock) and interpret the Hamiltonian constraint as an equation describing the evolution of the other degrees of freedom with respect to this chosen time variable.
This issue actually goes beyond the ``problem of time": due to the invariance of the theory under space-time diffeomorphism, it is impossible to localize space-time points using an a priori coordinate system. Coordinate systems have to be reconstructed a posteriori from the correlations between the various variables using internal degrees of freedom. This problem is usually addressed by re-introducing matter coupled to gravity and using it to localize space points and to define an internal clock with respect to which we can describe the evolution of the other degrees of freedom (see e.g. the work by Giesel and collaborators \cite{tina}).

The second issue is much harder to address. It is to construct an actual consistent quantum Hamiltonian constraint operator. One must regularize and quantize the classical Hamiltonian constraint, recognize the various ambiguities in the procedure, identify the ``correct" operator and check that the quantization is self-consistent and has the proper semi-classical regime (i.e. recover general relativity in a low-energy/large-scale regime). Since the initial proposal for a regularized Hamiltonian constraint operator by Thiemann \cite{thomas_dyn}, there haven't been a clear proposal to test a Hamiltonian operator or extract the semi-classical dynamics that it induces. In this context, one can not conclude the LQG quantization programme and define the physical Hilbert space $\cH_{phys}$ of quantum geometry states solving all the constraints of the theory. Of course, one should also wonder about the actual straightforward purpose of identifying the whole space of exact solutions to all the constraints and characterize explicitly the physical space $\cH_{phys}$: indeed, this would mean solving exactly quantum gravity, which sounds over-shooting since we don't know yet how to solve exact classical general relativity and characterize its whole space of classical solutions. It would seem more appropriate and more physically relevant to develop an approximation scheme to compute physical states and a perturbation expansion allowing to extract physical correlations and predictions from it. This should in particular allow to bridge between the non-perturbative background-independent LQG framework and the standard perturbative approach to quantum general relativity as a field theory around the flat space-time metric.

Addressing the problem of defining the dynamics of LQG had lead to several approaches, among which the {\it spinfoam programme} which define the Hamiltonian constraint operator through a path integral formalism, the master constraint programme (e.g. \cite{master}), the recent algebraic quantum gravity framework \cite{algqg}, the loop quantum cosmology framework which attempts to export the LQG quantization tools to cosmological models (e.g. \cite{lqc}), and the study of simplified LQG models based on restricted sets of spin network states based on simple graph and allowing to simulate simple physical settings \cite{lqg_models}.

\subsection{Covariant Loop Gravity and Projected Spin Networks}
\label{irrep}

To finish this overview of the canonical framework of loop quantum gravity, before moving on to spinfoam path integral formalism which implements the LQG dynamics, I would like to make a small detour through the framework of {\it covariant loop quantum gravity} \cite{clg,clqg}. Indeed it is a key step in relating the standard loop quantum gravity formalism as reviewed above to the usual spinfoam models constructed as discretization of general relativity formulated as a constrained BF theory. More precisely, the main discrepancy is that the standard formulation of spinfoam models is based on using the full Lorentz connection $\om$ while loop quantum gravity works in the time gauge where the time normal (to the canonical hypersurface $\Sigma$) is frozen to $(1,0,0,0)$ which reduces the connection to the Ashtekar-Barbero $\SU(2)$ connection. Covariant loop gravity was specifically introduced to defrost the time gauge and perform the canonical analysis of general relativity using the full original Lorentz connection $\om$ \cite{clg}. The canonical variables of covariant loop gravity are thus a Lorentz connection $\om$, its conjugate $\sl(2,\C)$-valued triad field $E$ and the time normal field $\chi\in\SO(3,1)/\SO(3)$ taking values in the upper hyperboloid $\cH^+$ of time-like unit vectors in the flat 3+1d Minkowsi space. I will not review the details of the canonical analysis in term of those variables, the interested reader will find all the details in \cite{clg}. Here, I would like to review the {\it projected spin network states}, introduced in \cite{projspinnet}, which are used to quantize the phase space of covariant loop gravity and which will turn out to be, on the one hand, the boundary states for spinfoam models and, on the other hand, very easily related to the $\SU(2)$ spin networks of loop quantum gravity.

Following the logic of loop quantum gravity with wave-functions depending on the $\SU(2)$ connection, the projective spin networks are wave-functions depending on the Lorentz connection and the time normal field \cite{projspinnet}. More precisely, we choose a graph $\Gamma$ and we define our wave-functions as cylindrical functions depending on the holonomies $U_e[\om]\in\SL(2,\C)$ of the Lorentz connection along the edges $e$ of the graph and on the values $\chi_v\in\SL(2,\C)/\SU(2)\sim\cH^+$ of the time normal field at the vertices $v$:
\be
\psi(\om,\chi)=\psi_\Gamma(\{U_e[\om],\chi_v\}).
\ee
These states are now Lorentz-invariant:
\be
\psi_\Gamma(\{U_e,\chi_v\})=
\psi_\Gamma(\{H_{s(e)}U_eH_{t(e)}^{-1},H_v.\chi_v\})
,\qquad\forall H_v\in\SL(2,\C),
\ee
but satisfy only a $\SU(2)$ gauge invariance if we freeze the time normal:
\be
\psi_\Gamma(\{U_e,\chi_v=\chi^{(0)}\})=
\psi_\Gamma(\{h_{s(e)}U_eh_{t(e)}^{-1},\chi^{(0)}\}),\qquad\forall h_v\in\SU(2),
\ee
for the time normal field fixed to $\chi^{(0)}=(1,0,0,0)\in\cH^+$.

Following \cite{projspinnet}, we use the Haar measure on $\SL(2,\C)$ to define the scalar product:
\be
\la\psi_\Gamma|\tpsi_\Gamma\ra\,=\,
\int_{\SL(2,\C)^E}\prod_edG_e \,\overline{\psi}(\{G_e,\chi_v\})\,\tpsi(\{G_e,\chi_v\}),
\ee
which can be easily seen not to depend on the values of the variables $\chi_v$ due to the $\SL(2,\C)$-invariance satisfied by those functionals. Thus the resulting Hilbert space is:
\be
L^2(\SL(2,\C)^E\times (\SL(2,\C)/\SU(2))^V/\SL(2,\C)^V)
\sim
L^2(\SL(2,\C)^E /\SU(2)^V).
\ee
Then, we can construct an orthogonal basis of this Hilbert space  using the Plancherel decomposition for $L^2$-function over  $\SL(2,\C)$, which is the extension of the Peter-Weyl theorem to non-compact Lie groups,
\be
f(G)=
\sum_n\int (n^2+\rho^2) d\rho\, \tr\,\left[F(n,\rho)D^{(n,\rho)}(G)\right],
\ee
where the couple $(n,\rho)$ labels irreducible representations (of the principal series) of $\SL(2,\C)$, the (infinite-dimensional) matrix $D^{(n,\rho)}(G)$ represents the group element $G\in\SL(2,\C)$ in the  $(n,\rho)$-representation and the matrix $F(n,\rho)$ gives the Fourier components of the initial function $f(G)$. We will call $R^{(n,\rho)}$ the Hilbert space attached to the  $(n,\rho)$-representation of $\SL(2,\C)$. It is interesting to decompose this representation space in term of representations of the $\SU(2)$ subgroup of $\SL(2,\C)$:
\be
R^{(n,\rho)}=\bigoplus_{j\in n+\N} V^j.
\ee
Finally, we give the values of the two Casimir operators of $\SL(2,\C)$ on $R^{(n,\rho)}$:
\be
C_1=\vec{K}^2-\vJ^2=\rho^2-n^2+1,
\qquad
C_2= \vJ \cdot \vec{K}=2n \rho,
\ee
where the $\vJ$ are the $\su(2)$ generators of the $\SU(2)$ subgroup while the $\vK$ are the boosts generators, $\sl(2,\C)=\su(2)\oplus \vK\C$. The reader can find more details on these representations in e.g.\cite{projspinnet,projspinnet2}. We apply this Plancherel  decomposition to our gauge-invariant functionals and build the projected spin network basis:
\bes
\psi^{\cI_e,j_e^{s,t},i_v}_\Gamma(G_e,\chi_v)
&\equiv\,
\sum_{m_e^s,m_e^t}&
\prod_e \la \cI_e,j_e^s,m_e^s|B_{s(e)}^{-1}G_e B_{t(e)}| \cI_e,j_e^t,m_e^t\ra\\
&&\prod_v \la \otimes_{e|t(e)=v}\,\cI_e,j_e^t,m_e^t| i_v |\otimes_{e|s(e)=v}\cI_e,j_e^s,m_e^s\ra.\nn
\ees
A projected spin network wave-function is labeled by $\SL(2,\C)$ representations $\cI_e=(n_e,\rho_e)$ on each edge $e$ of the graph, a couple of $\SU(2)$ representations with spin $j_e^{s,t}$ for each edge $e$, and finally a $\SU(2)$ intertwiner $i_v$ at each vertex. The last ingredient of the expression above is the group elements $B_v\in\SL(2,\C)$ which define the time normals $\chi_v$ from the origin $\chi^{(0)}=(1,0,0,0)$:
\be
\chi_v=B_v\,\chi^{(0)},\qquad B_v\in\SL(2,\C)/\SU(2).
\ee

\medskip

Now that we have reviewed the technical definition of these projected spin networks, let us discuss their relevance to linking loop quantum gravity and spinfoam models. First, let us remark that the projected spin networks satisfy as expected solely a $\SU(2)$ gauge invariance as soon as the time normal field is frozen i.e when the time gauge is imposed. This allows a direct link  between them and the standard $\SU(2)$ spin networks. Indeed if considering a projected spin network and evaluating it solely on the $\SU(2)$ subgroup, $\psi_\Gamma(\{g_e,\chi_v=\chi^{(0)}\})$ with all $g_e\in\SU(2)$, we recover simply a $\SU(2)$ cylindrical  function which decomposes itself onto spin network states. Indeed we can see that the structure of projected spin networks and $\SU(2)$ spin networks are very similar, and they both are constructed from $\SU(2)$ intertwiners. However, projected spin networks depend on much more data, a $\SL(2,\C)$ representation $\cI_e=(n_e,\rho_e)$ and an extra spin for  each edge $e$ of the graph. The projection map, consisting in evaluating the projected spin networks on the $\SU(2)$ subgroup, actually erases all this extra data. Details of the properties of this projection map and its inverse raising of $\SU(2)$ spin networks into projected spin networks can be found in \cite{projspinnet2}. All this to say that loop quantum gravity can be formulated in term of those projected spin networks, since their Hilbert space is isomorphic to the standard space of $\SU(2)$ spin networks up to a few subtleties (see \cite{projspinnet2} for a longer discussion).

Second, projected spin networks are the natural boundary states for the spinfoam models of the Barrett-Crane type or the EPRL-FK family. Indeed these models are constructed using the Lorentz connection. Moreover the time normal field $\chi$ is instrumental in implementing the {\it simplicity constraints} used to map the phase space of topological BF theory to the phase space of general relativity, which is the basic procedure to derive those spinfoam models.

\medskip

We have finished reviewing the main tools of loop quantum gravity. We will now describe the main features of the spinfoam framework, which allows to describe the dynamics of loop quantum gravity through evolving spin networks.

\section{2-Complexes and the Spinfoam Partition Function}

Spinfoams are originally histories of spin networks encoding the dynamics and evolution of quantum states for loop quantum gravity \cite{rr1}. The transition amplitudes associated to those histories have then been constructed mainly from the discretized path integral for topological BF theories (e.g. \cite{sfbaez}). Then, due mainly to the geometric interpretation of spin network states as discrete geometries, there appeared a relation between those spinfoam amplitudes and  Regge calculus for discretized general relativity, both in the large scale asymptotic limit and at the fundamental level. Finally, it should be recognized that, even if spinfoams arised from loop quantum gravity, the spinfoam framework for a quantum gravity  path integral is much more general that an attempt to derive the physical Hilbert space for loop quantum gravity and goes much beyond this original objective.

\subsection{Spin Network Histories}

The goal is to study the action of the Hamiltonian constraint operator on spin network states. This operator should generate the evolution of the spin networks in time. We formalize this dynamics in term of histories of evolving spin networks. Considering a spin network state based a graph $\Gamma$ and an arbitrary gauge group $\cG$, which is usually either $\SU(2)$ or the Lorentz group $\SL(2,\C)$, we have the graph $\Gamma$ dressed with $\cG$-representations on its links $l$ and with $\cG$-intertwiners on its nodes $v$. Let us now imagine this spin network states evolving in an extra-dimension (time). The links $l$ of the graph will sweep faces, to which are attached $\cG$-representations, and the nodes $v$ will sweep edges, on which live $\cG$-intertwiners. This creates a (combinatorial) 2-complex with edges and faces, which describes the evolving spin network states. The vertices of this 2-complex then represent space-time points, where the graph $\Gamma$ changes. At these points, an edge splits in several edges, or vice-versa, leading to new faces and to a different graph. This 2-complex structure is called a {\it spinfoam}:

\begin{center}
\begin{tabular}{lcl}
spin network & $\arr$ & spinfoam \\
representations $j_l$ on links $l$ & $\arr$ & representations $j_f$ on faces $f$ \\
intertwiners $i_v$ on nodes $v$ & $\arr$ & intertwiners $i_e$ on edges $e$ \\
changes of graph $\Gamma$ & $\arr$ & spinfoam vertices $\sigma$ encoding the dynamics
\end{tabular}
\end{center}

\noindent
Reciprocally, if we take a slice of the spinfoam 2-complex, we get a spin network. From this viewpoint, spin network states are the boundary states for spinfoams.

Considering an arbitrary 2-complex, an important concept for describing and understanding spinfoams is the boundary spin network states around each vertex $\sigma$. Let us focus on a single spinfoam vertex $\sigma$ and let us consider a small sphere around it which contains only this spinfoam vertex and no other. Around $\sigma$, there are spinfoam edges $e$  which leave from $\sigma$ and spinfoam faces $f$ which are between couples of edges. These edges and faces intersect the small sphere bounding the vertex $\sigma$ and create a spin network state: the edges puncture the sphere at nodes $v$ and the intersection of the faces with the sphere gives the links $l$ relating those nodes. The links inherit the representations living on the corresponding faces and the nodes inherit the intertwiners living on the corresponding edges. This is called the {\it boundary spin network} $\psi_\sigma$ dual to the vertex $\sigma$.

Then we can view the whole spinfoam 2-complex as the set of those little bubbles around each vertex $\sigma$ and on which live the corresponding boundary spin network state. This picture of the quantum space-time made from those little bubbles with spin network actually lead to the name ``spinfoam" for this framework.

\medskip

To conclude this initial presentation of the basic structure of spinfoams, we need to point out that the usual 2-complexes that are used in spinfoam models are the dual 2-skeleton to space-time triangulations. This equivalence was realized right from the beginning of the development of the spinfoam framework e.g. \cite{fotini}.

Starting with the simpler case of a three-dimensional space-time, a space-time triangulation consist in tetrahedra glued together along their triangles. The  dual 2-skeleton is defined as follows. The spinfoam vertices $\sigma$ are dual to each tetrahedron. Those vertices are all 4-valent with the four attached edges being dual to the four triangles of the tetrahedron. Each edge $e$ then relates two spinfoam vertices, representing the triangle which glues the two corresponding tetrahedra. Finally, the spinfoam faces $f$ are reconstructed as dual to the triangulation's edges. Indeed, considering an edge of the triangulation, we go all around the edge and look at the closed sequences of spinfoam vertices and edges which represent respectively all the tetrahedra and triangles that share that given edge. This line bounds the spinfoam face, or {\it plaquette}, dual to that edge. Finally, each spinfoam  edge $e$ has three plaquettes around it, representing the three triangulations edges of its dual triangle. To summarize the situation:

\begin{center}
\begin{tabular}{lcl}
3d triangulation & $\leftrightarrow$ & spinfoam 2-complex\\\hline
tetrahedron $T$ &  & 4-valent vertex $\sigma$ \\
triangle $t$ &  &  edge $e$ \\
edge &  & plaquette $f$
\end{tabular}
\end{center}

The setting is very similar for the four-dimensional case. The triangulated space-time is made from 4-simplices glued together at tetrahedra. Each 4-simplex is a combinatorial structure made of 5 boundary tetrahedra, glued to each other through 10 triangles. Once again, we define the spinfoam 2-complex as the dual 2-skeleton:

\begin{center}
\begin{tabular}{lcl}
4d triangulation & $\leftrightarrow$ & spinfoam 2-complex\\\hline
4-simplex & & spinfoam vertex $\sigma$ \\
tetrahedron $T$ &  & spinfoam edge $e$ \\
triangle $t$ &  &  plaquette $f$
\end{tabular}
\end{center}

\noindent
The two other levels, with the triangulation's edges and points, are ignored by the dual 2-skeleton. However, when looking at a given spinfoam and its associated 2-complex and considering its geometric interpretation as an actual discrete space-time geometry, these space-time points and triangulation's edges will have to be reconstructed a posteriori. This reconstruction of these two levels actually leads to pseudo-triangulations with conical singularities and other surprises (see e.g. \cite{pseudotriangl}).

Finally, this correspondence between space-time triangulations and spinfoams is easily generalized to arbitrary space-time dimension and to arbitrary cellular decomposition:

\begin{center}
\begin{tabular}{lcl}
$n$-d triangulation & $\leftrightarrow$ & spinfoam 2-complex\\\hline
$n$-d simplex & & spinfoam vertex $\sigma$ \\
$(n-1)$-d cells &  & spinfoam edge $e$ \\
$(n-2)$-d cells &  &  plaquette $f$
\end{tabular}
\end{center}

\noindent
Around each of the spinfoam vertex $\sigma$, we can reconstruct its boundary spin network whose graph is dual to the $(n-1)$-d cells and $(n-2)$-d cells contained by the corresponding $n$-d simplex.

\subsection{The Local Spinfoam Ansatz}

A spinfoam model is then defined as a choice of the set of 2-complexes which will be considered and a choice of probability amplitude for each possible spin network history i.e for each 2-complex dressed with representations and intertwiners.

The local spinfoam ansatz is to construct the amplitude associated to a 2-complex from the product of local amplitudes associated to the vertices, edges and faces of the spinfoam and only depending on the local representations and intertwiners living on those cells. More precisely, let us consider a given 2-complex $\Delta$ with boundary $\pp\Delta$ and a spin network state which defines the boundary state on $\pp\Delta$. Then the spinfoam amplitude associated to $\Delta$ consists in a sum over all possible representations and intertwiners living in the bulk and consistent with the boundary spin network:
\be
\cA[\Delta,\psi_{\pp\Delta}]
\,=\,
\sum_{j_f,i_e} \prod_f \cA_f[j_f]\,\prod_e \cA_e[i_e,j_{f\ni e}]
\,\prod_\sigma \cA_\sigma[j_{f\ni\sigma},i_{e\ni\sigma}],
\ee
where the representations and intertwiners $j_f,i_e$ for faces and edges on the boundary $f,e\in\pp\Delta$ are fixed and given by our choice of boundary state $\psi_{\pp\Delta}$. This is the local ansatz for spinfoam amplitudes.

The amplitude $\cA[\Delta,\psi_{\pp\Delta}]$ defines our discrete spinfoam path integral. The face weight $\cA_f$ and edge weight $\cA_e$ are considered as kinematical and defining the path integral measure, while the vertex amplitude $\cA_\sigma$ contains all the dynamical information of the spinfoam model.

The standard choices of amplitudes for the usual spinfoam models are the face weight $\cA_f[j_f]$ given by the dimension of the representation $\dim j_f$ and the vertex amplitude $\cA_\sigma$ given by the evaluation of the boundary spin network for $\sigma$. This means evaluating this boundary spin network functionals on the trivial holonomies i.e fixed to the identity. It gives simply the straightforward trace of the intertwiners $i_{e\ni\sigma}$ contracted together along the combinatorial structure of the boundary graph. Finally, the edge weight $\cA_e$ can be totally re-absorbed in the vertex amplitude.

If we assume this standard choice for the local amplitudes, then the freedom in defining a spinfoam model is very restricted. Indeed, the spinfoam model will be entirely determined by, first, the choice of the gauge group $\cG$, second, the choice of possible representations and intertwiners labeling the faces and edges. Both the Barrett-Crane models and the EPRL-FK family of spinfoam models are of the type and are characterized by different choices of representations and intertwiners.

\medskip

Finally the last step of the spinfoam programme is to get rid of the bulk triangulation, that is more precisely sum over all possible 2-complexes compatible with the boundary data:
\be
\cA[\Gamma,\psi_\Gamma]
=\sum_{\Delta|\pp\Delta=\Gamma}w(\Delta)\,\cA[\Delta,\psi_\Gamma],
\ee
where $w(\Delta)$ is a statistical weight depending only on the 2-complex $\Delta$ (and not on the representations and intertwiners living on it and defining the spinfoam), which usually simply depends on its symmetries. This sum is much less controled by the previous sum over representations and intertwiners for a fixed 2-complex. It is usually ill-defined, it naively diverges and requires proper gauge-fixing. The standard way to define it non-perturbatively is through the group field theory formalism, which uses generalized matrix models to generate the sum over suitable 2-complexes, as we will review later.

Considering this final expression for the spinfoam amplitude, we will remark that if we choose a disconnected boundary $\pp\Delta=\Gamma=\Gamma_{in}\cup\Gamma_{out}$, then this spinfoam amplitude is interpreted as a transition amplitude between the initial state $\psi_{\Gamma_{in}}$ and the final state $\psi_{\Gamma_{out}}$. Now the goal is to construct suitable consistent amplitudes $\cA_f,\cA_e,\cA_\sigma$ in order that the spinfoam amplitude $\cA[\Delta]$ correctly reproduces the dynamics generated by the Hamiltonian constraint for loop quantum gravity.

\section{Constructing Spinfoam Amplitudes from Topological BF Theory}

Now that we have described the general structure of spinfoams, we need to construct the explicit probability amplitudes. For this purpose, we use a reformulation of general relativity as a constrained BF theory. Topological BF theory is indeed a very convenient start point to build spinfoam amplitudes. First, it is a topological field theory, it does not have local degrees of freedom and can be discretized without affecting the physical content of the theory. Second, we know how to discretize and build its path integral on space-time triangulations into spinfoam models. The logic is then to build the spinfoam path integral for general relativity by perturbing around the BF path integral and deforming it. The first step of this programme is to understand how to write gravity as a BF theory at the classical level, then to understand which constrains transform BF theory into general relativity and see how impose them on the spinfoam path integral at the quantum level. This is what we will review in this section.

\subsection{General Relativity as a Constrained Topological BF Theory}

Topological BF theory in four space-time dimension for a gauge group $\cG$ has a simple action:
\be
S[\om,B]\,=\,
\int_\cM \tr B\,\w F[\om].
\label{bf}
\ee
The basic variables are a $\ag$-valued connection $\om$ and a $\ag$-valued 2-form $B$, where $\ag$ is the Lie algebra for the gauge group $\cG$. The classical equations of motion impose that the connection  is flat, $F[\om]=0$, and that the $B$-field has a trivial parallel transport, $d_\om B=0$. This theory is topological and does not have any local degree of freedom, which can be confirmed by a straightforward counting of the degrees of freedom in a canonical analysis. Such a theory is easy to study at the quantum level, due to the absence of local degrees of freedom and to the flatness of the connection. This is the starting point for the spinfoam quantization programme.

Simply comparing the BF action above with the Holst-Palatini action for general relativity \Ref{holst}, at the first glance, the two actions are very similar: gravity is a BF theory for the gauge group $\cG=\SL(2,\C)$ and with the $B$-field related to the tetrad field $B=\star(e\w e)+\f1\gamma e\w e$. More precisely, following \cite{BFimm1,BFimm2}, we introduce the modified BF action, also called generalized Plebanski action:
\be
S=\int B^{IJ}\w F_{IJ} -\f{1}{2}\phi_{IJKL}B^{IJ}\w B^{KL}+\mu C,
\ee
where $C=a_1\phi_{IJ}\,^{IJ}+a_2\phi_{IJKL}\epsilon^{IJKL}$ with $a_{1}$ and $a_{2}$ two arbitrary constants. The scalar field  $\phi$  is a Lagrange multiplier with the symmetries $\phi_{IJKL}=-\phi_{JIKL}=-\phi_{IJLK}=\phi_{KLIJ}$. Finally, the 4-form $\mu$ is a Lagrange
multiplier enforcing the condition $C(\phi)=0$ on the field $\phi$.

It is clear that the field $\phi$ imposes some constraints on the $B$-field. Since $B$ is itself the Lagrange multiplier enforcing the flatness of the connection $\om$, the new constraints on $B$ actually relax this condition of the flatness of the connection and allow to re-introduce local degrees of freedom into the BF theory (corresponding to the local fluctuations of the curvature $F[\om]$). More precisely, taking into account the symmetries of $\phi$ and the constraints $C=0$, we can show that the constraints on the field $B$ read:
\be
B^{IJ}\w B^{KL} =\f{1}{6} (B^{MN}\w B_{MN}) \eta^{[I |K|} \eta^{J]L}
-\f{1}{12}(B^{MN}\w \star B_{MN}) \epsilon^{IJKL},
\ee
$$
2a_2 B^{IJ}\w B_{IJ} -\epsilon a_1 B^{IJ}\w \star B_{IJ}=0,
$$
where $\star$ s the Hodge operator acting on internal Lorentz indices,
$\star B_{IJ}=1/2\,\epsilon_{IJKL}B^{KL}$ with $\star^2=-1$ (or  $\star^2=+1$
if working with a Euclidean signature). These are called the {\it simplicity constraints} on the $B$-field. We can prove that they imply that the $B$-field can be entirely reconstructed in term of a tetrad field \cite{fdp,BFimm1,BFimm2}:
\be
B^{IJ}=\alpha *(e^I \w e^J) + \beta\, e^I \w e^J,
\qquad\textrm{with}\quad
\f{a_2}{a_1}=\f{\alpha^2-\beta^2}{4\alpha\beta}.
\ee
Thus the simplicity constraints mean that the field $B$ determines the whole geometry: one can reconstruct the tetrad and the metric from it. They turn the topological BF theory into a theory of the space-time geometry. Re-injecting the expression above in the BF action, we recover general relativity for an Immirzi parameter $\gamma=\f\alpha\beta$:
$$
S=\alpha\,\left( \int  \star(e^I \w e^J)\w F_{IJ}  +
\f1\gamma \int e^I \w e^J \w F_{IJ}\right).
$$
Expressed in term of $\gamma$, the condition on the two coupling constants $a_{1,2}$ read:
$$
\f{a_2}{a_1}=\f{1}{4}\left(\gamma -\f{1}{\gamma}\right).
$$
For the interested reader, all this formalism can also be written in term of the self-dual and anti-self-dual $\SU(2)$ connections \cite{actionfk,BFimm2,michael}.

The big subtlety about this approach of writing general relativity as a constrained BF theory is the existence of different geometric sector. Indeed , the modified BF action is invariant under the discrete transformation $B\arr\star B$. This switches the role of the constants $\alpha$ and $\beta$, thus mapping the Immirzi parameter $\gamma$ to its inverse. More precisely, there exists four sectors obtained by successive applications of the Hodge dual operator:
\be
(\alpha,\beta) \,\arr\, (\beta,-\alpha) \,\arr\,
(-\alpha,-\beta) \,\arr\, (-\beta,\alpha)\,.
\ee
These four sectors have to be distinguished when constructing the spinfoam amplitude. This is usually done by a careful analysis of the properties of the spin vertex amplitude e.g. \cite{fk}.
A step forward understanding the mechanism seperating these sectors at the classical level is the re-writing of the constrained BF action with linear constraints on the $B$-field \cite{linearB}.

\medskip

To conclude this classical presentation, I would like to make three remarks. The first one is about the original Plebanski action. It corresponds to the case without Immirzi parameter, i.e here $a_1=0$. The four sectors of the BF theory correspond to the cases of a vanishing or infinite Immirzi parameter $\gamma$:

\begin{center}
\begin{tabular}{lcl}
sector (I$\pm$) &$\quad$& $B=\pm\star(e\w e)$, \\
sector (II$\pm$) &$\quad$& $B=\pm(e\w e)$.
\end{tabular}
\end{center}

\noindent
The sectors (I$\pm$) give back general relativity while the sectors (II$\pm$) are actually non-geometric and don't impose a vanishing torsion. It is therefore crucial to distinguish these sectors in the spinfoam construction and get rid of the contributions of the non-geometric sectors (II$\pm$) in the path integral.

The second remark is that the four regimes actually co\"incide when the tetrad field $e$ (or equivalently the $B$-field) is degenerate: it is the degenerate metrics which allow to navigate between these four regimes in the path integral.

The third and last point is that there exists a different type of modified BF action for general relativity. This is based on the MacDowell-Mansouri action, revisited by Freidel-Starodubtsev, which uses a $\Spin(4,1)$-action instead of our Lorentz connection $\om$, and a symmetry breaking interaction term to reduce the gauge symmetry from $\Spin(4,1)$ down to $\Spin(3,1)$ as wanted \cite{mcdm}. Unfortunately, this very interesting re-formulation has not lead yet to the construction of explicit spinfoam models.

\subsection{Discretizing the Path Integral}

Having rewritten general relativity as a constrained BF theory, we will write the discretized path integral for BF theory as a spinfoam model and see how to twist it and deform it to obtain a realistic proposal for a spinfoam path integral for quantum gravity.

Let us start for BF theory with gauge group $\SU(2)$ in three space-time dimensions. This actually is exactly 3d Euclidean gravity, which is a topological field theory. And its spinfoam quantization leads to the famous Ponzano-Regge state-sum model \cite{pr}. The classical action reads:
\be
S[B,\om]=\int_{\cM_{3d}} B^i\w F^i[\om],
\ee
where $\om$ is a $\SU(2)$ connection and $B$ a $\su(2)$-valued 1-form defining the triad. The $B$-field is a straightforward Lagrange multiplier enforcing the flatness of the $\SU(2)$ connection, $F[\om]=0$. Considering a cellular decomposition (or a triangulation for the sake of simplicity) of the 3d space-time and its dual 2-complex $\Delta$, the discretization procedure is simple. We discretize the connection on spinfoam edges and we discretize the $B$-field on edges of the cellular decomposition or equivalently on spinfoam faces, which we summarize in the following table.

\begin{center}
\begin{tabular}{l|l|l}
triangulation & spinfoam & discrete field \\ \hline
triangle & edge $e$ & $U_e[\om]\in\SU(2)$ \\
edge & face $f$ & $B_f\in\su(2)$ \\
\end{tabular}
\end{center}

The goal is to write the BF path integral,
$$
Z=\int [dB][d\om]\,e^{i\int_{\cM} B^i\w F^i[\om]} \,=\,
\int [d\om]\,\delta^{(3)}(F^i[\om]),
$$
in this discretized setting. Taking into account that the path integral reduces basically to the flatness condition, the natural proposal for the discrete path integral is:
\be
Z_\Delta\,=\,
\int_{\SU(2)} \prod_{e\in \Delta} dU_e\,\prod_{f\in\Delta}\delta\left(\overrightarrow{\prod}_{e\in \pp f} U_e\right).
\ee
The oriented product $g_f\,\equiv\,\overrightarrow{\prod}_{e\in \pp f} U_e$ around the plaquette $f$ defines the closed holonomy around that plaquette, or equivalently around the dual edge of the triangulation. Then the $\delta(\cdot)$-function are the standard $\delta$-distribution on the $\SU(2)$ group imposing that the group element is the identity. This discretized path integral clearly imposes that all holonomies are trivial, thus that the connection is flat.

This path integral is not yet written as a local spinfoam model, but it is actually one of the most compact and efficient way to write spinfoam path integrals. Indeed, the physical content of the model is entirely transparent written as such an integral. This is called the {\it connection representation} of spinfoam amplitudes. It is dual to the local spinfoam ansatz defined above in the sense that the spinfoam amplitude in the connection representation  are expressed in term of group elements while they are expressed in term of group representations in the local spinfoam ansatz.

Already at this level, we can make two important remarks. First, it is possible to (re-)introduce the discretized $B$-field in this path integral to write it in term of a discretized BF action principle \cite{actionfk,pr1}:
\be
Z_\Delta\,=\,
\int_{\SU(2)} \prod_{e\in \Delta} dU_e\,\int_{\su(2)}\prod_f dB_f\,
e^{\sum_f \tr B_f g_f},
\qquad\textrm{with}\quad
g_f\,=\,\overrightarrow{\prod}_{e\in \pp f} U_e.
\ee
The trace in the exponential is taken in the fundamental representation in which the algebra elements $B_f$ and group elements $U_e$ are given as 2$\times$2 matrices. There are nevertheless a few subtleties in the equivalence between this discrete action principle and the initial $\delta$-path integral \cite{pr1,Bobs}, mainly due to the difference between $\SU(2)$ and $\SO(3)$.

The second remark is that since the path integral is written as a product of $\delta$-distribution, it is inevitable that some of those $\delta$-functions are redundant and that the path integral formally defined as such diverges due to those redundancies \cite{pr1}. However, it has been shown that those redundancies and naive divergences are due to a symmetry of the discrete path integral, which can be identified as a invariance under discrete diffeomorphisms \cite{pr1,pr2}. Using a generalization of the tree techniques introduced to gauge-fix spin network funcionals \cite{noncompact}, it has also been shown that the redundancies can be completely gauged out and that the path integral can be defined with a minimal number of $\delta$-distribution \cite{pr1,pr2,pr3}.

The final step of the definition of the spinfoam model is to decompose the $\delta$-distribution over $\SU(2)$ representations using the Peter-Weyl theorem,
\be
\delta(g)=\sum_j d_j\,\tr D^j(g),
\ee
and to perform explicitly all the integrations over the $\SU(2)$ group elements $U_e$. Considering the special case when the cellular decomposition is simply a triangulation for simplicity's sake, the integrals over $\SU(2)$ are all of the following type:
\be
\int_{\SU(2)} dg\,
\prod_{i=1}^3 D^{j_i}_{a_ib_i}(g)
\,=\,
C^{j_1j_2j_3}_{a_1a_2a_3}C^{j_1j_2j_3}_{b_1b_2b_3},
\ee
where the $C$'s are the normalized Clebsh-Gordan coefficients (or more exactly the Wigner 3j-symbols) for the recoupling theory of $\SU(2)$ representations. This integral can actually derived and interpreted as the decomposition of the identity on the space of 3-valent intertwiners in term of the standard  Clebsh-Gordan basis. This is almost trivial in this 3-valent situation since there exists a single 3-valent intertwiners (up to a global factor) between three fixed spin $j_1,j_2,j_3$. If we relax the condition that the cellular decomposition is a triangulation, we obtain higher integrals which are then computed similarly as the decomposition of the identity on the space of $n$-valent intertwiners.
We have one of those integrals for each integration over a group element $U_e$ living on an edge $e$, i.e for each triangle of the triangulation. Regrouping this Clebsh-Gordan coefficients appropriately according to the tetrahedron $T$ of the triangulation or equivalently to the spinfoam vertices $\sigma$, we can write our discrete path integral as:
\be
Z_\Delta\,=\,
\sum_{\{j_f\}}\prod_f d_{j_f}\,\prod_T \{6j\}_T,
\ee
where the vertex amplitude is given by the 6j-symbol for the recoupling theory of $\SU(2)$ representations:
\be
\{6j\}_T\,\equiv\,
\left\{
\begin{array}{ccc}
j^T_1 & j^T_2 &j^T_3\\
j^T_4 & j^T_5 &j^T_6
\end{array}
\right\}
\,=\,
\sum_{\{a_i\}}
C^{j_1j_2j_3}_{a_1a_2a_3}C^{j_3j_4j_5}_{a_3a_4a_5}C^{j_5j_2j_6}_{a_5a_2a_6}C^{j_6j_4j_1}_{a_6a_4a_1},
\ee
where the six spins $j^T_i$ are the six representations living on the edges of the tetrahedron on the triangulation. This vertex amplitude $\{6j\}_T$ is also the evaluation of the boundary spin network around the vertex $\sigma$ (dual to $T$). As we have said earlier, it is given by the straightforward contraction of the intertwiners living on this boundary spin network.

This final expression is the defining formula for the Ponzano-Regge spinfoam model \cite{pr}. It is written exactly under the correct form for a local spinfoam amplitude. It has been checked explicitly that this path integral is a topological state-sum invariant Pachner moves (due to the Biedenharn-Elliot identity) and that it defines as expected a projector onto flat $\SU(2)$ connection on the boundary e.g. \cite{pr_ooguri}.

\medskip

We can repeat the same steps to deal with the path integral for topological BF theory in four space-time dimensions. Working again with the $\SU(2)$ gauge group, we now have a $\SU(2)$-connection $\om$ and a $\su(2)$-valued 2-form $B$. We discretize the connection on the spinfoam edges $e$ and the $B$-field on the spinfoam faces $f$ or equivalently on the triangles $t$ of the dual triangulation.

\begin{center}
\begin{tabular}{l|l|l}
triangulation & spinfoam & discrete field \\ \hline
tetrahedron & edge $e$ & $U_e[\om]\in\SU(2)$ \\
triangle & face $f$ & $B_f\in\su(2)$ \\
\end{tabular}
\end{center}

\noindent
The path integral is then formally exactly as the same as before:
\be
Z_\Delta\,=\,
\int_{\SU(2)} \prod_{e\in \Delta} dU_e\,\prod_{f\in\Delta}\delta\left(\overrightarrow{\prod}_{e\in \pp f} U_e\right).
\ee
The difference with the 3d case is the combinatorial structure of the 2-complex, which is now dual to a 4d cellular decomposition. Indeed, looking at the integration over the group element $U_e$, the edge $e$ is dual to a tetrahedron $T$ which has four triangles: consequently the edge $e$ will be part of four faces $f$, and the integral over $U_e$ will consist in the product of four representation matrices. Nevertheless, as we already said earlier, such integral are straightforward to compute:
\be
\int_{\SU(2)} dg\,
\prod_{i=1}^4 D^{j_i}_{a_ib_i}(g)
\,=\,
\sum_J C^{j_1j_2j_3j_4\,J}_{a_1a_2a_3a_4}C^{j_1j_2j_3j_4\,J}_{b_1b_2b_3b_4},
\ee
where $J\in\N/2$ is an internal spin index labeling the basis of 4-valent intertwiners between the representations $j_1,j_2,j_3,j_4$. Those 4-valent intertwiners can be easily written in term of the usual 3-valent Clebsh-Gordan coefficients:
\be
C^{j_1j_2j_3j_4\,J}_{a_1a_2a_3a_4}\,=\,
\sum_a
C^{j_1j_2J}_{a_1a_2a}C^{j_3j_4J}_{a_3a_4a},
\ee
up to some normalization factors.
Using this formula to compute the discrete path integral, after re-grouping appropriately the various terms, we finally obtain:
\be
Z_\Delta\,=\,
\sum_{\{j_f,i_e\}}\prod_f d_{j_f}\,\prod_e \Xi(i_e)\,
\prod_\sigma \{15j\}_\sigma.
\ee
Here, we have attributed a representation $j_f$ to each spinfoam face or equivalently to each triangle $t$. The corresponding face weight $d_{j_f}$ is simply the dimension of that representation. Each edge $e$ is dual to a tetrahedron $T$. According to the integration formula above, we have attached one spin $J_e$ to each of the edges/tetrahedra, which actually defines a 4-valent intertwiner $i_e$ between the four representations living on the triangles of the tetrahedron $T$ (or equivalently living on the four faces to which the edge $e$ belongs). The symbol $\Xi(i_e)$ is simply the norm of the intertwiner, which we can set to 1.

Finally, looking at a spinfoam vertex $\sigma$, or equivalently to its dual 4-simplex, it belongs to five spinfoam edges (or tetrahedra) and to ten spinfoam faces (or triangles). The corresponding boundary spin network lives on the now famous 4-simplex graph with 5 vertices (corresponding to the tetrahedra) and 10 links between them (corresponding to the triangles). The evaluation of this boundary spin network, labeled by 10 representations $j_f$ and 5 intertwiners $i_e$, is defined as the straightforward contraction to the five intertwiners together and gives the 15j-symbol of the recoupling theory of $\SU(2)$ representations.

\medskip

This concludes this overview of the construction of the spinfoam path integral for the topological BF theory. We have dealt with the $\SU(2)$ gauge group up to now. To study Euclidean 4d gravity, we will have to use as gauge group $\Spin(4)=\SU(2)_L\times \SU(2)_R$ which leads very simply to two (decoupled) copies of the $\SU(2)$ BF path integral. As for the more relevant case of Lorentzian 4d gravity, we use as gauge group the non-compact Lie group $\SL(2,\C)$. All the procedure is exactly the same, but we now use $\SL(2,\C)$-representations and $\SL(2,\C)$-intertwiners on the faces and edges of the spinfoam. Using a non-compact group leads to a few divergences when performing the integrations over the group. However, all these of can easily controlled and gauged-away.

\medskip

We have seen here how to write the spinfoam amplitudes, either in term of integrals over group elements or in term of sums over group representations. These two equivalent way of defining the spinfoam amplitudes have each their advantages and inconvenients. The connection representation is particularly efficient when investigating the (gauge) symmetries of the path integral and its relation to BF theory. The expression of the amplitudes in term of representations and 3$nj$-symbols is useful when computing numerically the amplitudes and when studying their large scale asymptotics (for large spins).

\subsection{Simplicity Constraints and Spinfoam Models for 4d Gravity}

So we have, on the one hand, reformulated general relativity as a BF theory with gauge group $\SL(2,\C)$ with constraints and, on the other hand, shown how to write the path integral for BF theory as a spinfoam model. The natural next step is to investigate how to implement the constraints in the path integral and how to modify the BF spinfoam model to derive a proposal for a quantum gravity path integral.

The standard procedure is to discretize the simplicity constraints, which enforce that the $B$-field comes from a tetrad field $e$ and thus have a geometric meaning. The starting point is always the same, we discretize the $B$-field on spinfoam faces (actually more precisely, we discretize it on spinfoam wedges, which are obtained by cutting the faces into smaller pieces which each contain a single spinfoam vertex) and the Lorentz connection $\om$ as holonomies living on the spinfoam edges. Then they are two ways to proceed:
\begin{itemize}

\item Either, we turn the simplicity constraints as quantum operators. We apply them directly at the quantum level on the intertwiners and boundary spin network states (around the spinfoam vertex). We solve them and define the resulting spinfoam model, which depends on the specific quantum simplicity constraints and the particular solutions that we select. For instance, we are usually able to distinguish the solutions that correspond to the various classical sectors of the constrainted BF theory and identify the ones relevant to the sector that we want to spinfoam quantize.

    This way of proceeding is similar to a {\it geometric quantization}. Indeed, first, the discretized $B_f$-variables are attached to the spinfoam faces or equivalently to their dual triangles. They are geometrically identified as the normal bivectors to the triangles $t$. Then, one usually focus on the simplicity constraint operators within each 4-simplex, i.e as acting on the boundary spin network around each spinfoam vertex $\sigma$. This boundary spin network is interpreted as defining the quantum state of the  4-simplex with normal bivectors given by the $B_f$'s, and the simplicity constraints are interpreted as ensuring that the 4-simplex is geometric, i.e that it can be embedded in the flat Minkowski space-time. Finally, one must investigate how to glue back the 4-simplices in a geometrically consistent way.

\item Or the second method is to discretize the simplicity constraints and include them at the classical level in a discrete constrained BF action. This provides us with a connection representation of the spinfoam amplitude and then we can compute the corresponding discrete path integral. Compared to the first method, the logic of this procedure is clearer and the relation of the derived spinfoam model with the original BF theory is also more transparent. Furthermore, we do not have to worry about the gluing of 4-simplices together, since this is already taken into account in the discrete BF action from which we started.

    Finally, although this method seems cleaner, it usually leads to spinfoam amplitudes much more complicated than the ones obtained from the first method. Moreover, it very often leads to spinfoam amplitudes which are not local. Nevertheless, it seems that this point of view allows an easier link with Regge calculus and could be more useful than the previous one when studying the coarse-graining and renormalization of the spinfoam amplitudes.

\end{itemize}

I will not review the various proposals of discretized simplicity constraints and the resulting procedure to quantize them and solve them. There have been indeed  many proposals, which finally seem to converge towards the same geometrical picture, although the details of the resulting spinfoam amplitudes can be subtly different. Here is a surely not-complete list of the work on this topic: the first proposals of discretizing and quantizing the simplicity constraints \cite{michael,actionfk,puzio}, the geometric quantization \`a la Barrett-Crane by imposing strongly the simplicity constraints on every 4-simplex \cite{bc1,bc2,reis_vertex}, proposals to impose the simplicity constraints in a weaker sense \cite{ls1,epr}, proposals to reformulate the simplicity constraints in term of the time normal \cite{ls2,eprl,sf_sergei} and the resulting last generation of spinfoam models of the EPRL-FK family \cite{fk,eprl}, and finally possible ways to write them in a discrete action principle \cite{actionfk,actionfc,actionlb}.

\medskip

Instead of going into the details of deriving specific spinfoam models, I will just give the final resulting amplitudes for the most studied spinfoam models - the Barrett-Crane model and the EPRL model.

So starting for topological BF theory for the Lorentz group, the partition function on a given 2-complex $\Delta$ reads in the connection representation:
\be
Z_\Delta\,=\,
\int_{\SL(2,\C)} \prod_{e\in \Delta} dG_e\,\prod_{f\in\Delta}\delta\left(\overrightarrow{\prod}_{e\in \pp f} G_e\right).
\ee
The Barrett-Crane model is a spinfoam model for 4d quantum gravity without Immirzi parameter. It is given by  a slight modification of the BF functional:
\be
Z^{BC}_\Delta\,=\,
\int_{\SL(2,\C)} \prod_{e\in \Delta} dG^s_edG^t_e\,
\int_{\SU(2)}\prod_{f,e\in \pp f}dh_{e,f}\,
\prod_{f\in\Delta}\delta^{BC}\left(\overrightarrow{\prod}_{e\in \pp f} G^s_eh_{e,f}(G^t_e)^{-1}\right),
\ee
where the modified distribution $\delta^{BC}$ is defined over simple representations of $\SL(2,\C)$ of the type $(n=0,\rho)$:
\be
\delta^{BC}(G)\,=\,
\sum_\rho \rho^2\,\tr D^{(0,\rho)}(G),
\ee
where the measure $\rho^2$ is the Plancherel measure and defines the face weight $\cA_f(\rho_f)$ of the model. $D^{(0,\rho)}(G)$ is the matrix representing the group element $G$ in the irreducible $\SL(2,\C)$ representation labeled by $(n=0,\rho)$.

To make the connection with the more usual expression of the Barrett-Crane model written as a local spinfoam amplitude, one first consider a 2-complex dual to a 4d triangulation.
Then, if one expand the $\delta^{BC}$-distribution into $\SL(2,\C)$-representation and performs the integration over the $\SU(2)$ group elements, the unique Barrett-Crane intertwiner will appear on every edge, i.e every tetrahedron. Finally performing the integration over the $\SL(2,\C)$ group elements, one can regroup the terms according to which spinfoam vertex (or equivalent 4-simplex) they refer to and one can see the standard $\{10\rho\}$-symbol appearing for every 4-simplex:
\be
Z^{BC}_\Delta\,=\,
\sum_{\{\rho_f\}}\prod_f\rho_f^2\,
\prod_\sigma \{10\rho\}_\sigma,
\ee
with the symbol $\{10\rho\}_\sigma$ depending on the 10 representations $\rho_f$ living on the ten faces (or triangles) of the 4-simplex $\sigma$:
\be
\{10\rho\}_\sigma=\int \prod_{v=1}^5dG_v\,\prod_{l=1}^{10}K^{\rho_l}(G_{s(l)}G_{t(l)}^{-1}).
\ee
Here we have written the 10$j$-symbol as the evaluation of the boundary spin network around the spinfoam vertex $\sigma$: the vertices $v$ of the boundary spin network correspond to the 5 tetrahedra of $\sigma$ and the 10 links correspond to the 10 triangles of $\sigma$. The kernel $K^\rho$ is defined as the matrix element of the group element on the $\SU(2)$-invariant vector in the $(0,\rho)$-representation:
\be
K^\rho(G)\,\equiv\,
\la (0,\rho)\,, j=0\,|\,G\,|\,(0,\rho)\,, j=0\ra.
\ee
Indeed, as we saw earlier in section \ref{irrep}, the $\SL(2,\C)$ representation $(n=0,\rho)$ contains a unique $\SU(2)$ invariant vector given for $j=n=0$.

Defining the Barrett-Crane model in the connection representation has many advantages:
\begin{itemize}
\item It is defined for arbitrary 2-complexes, without referring to a dual space-time triangulation.
\item It allows to a closer relation with the discrete BF theory, especially when trying to write a discrete action principle for the Barrett-Crane model e.g. \cite{actionlb}.
\item It allows an easier reformulation as Feynman diagrams of a group field theory \cite{fdpkr,GFTbc}.
\end{itemize}

One can define the EPRL model in a similar fashion \cite{eprl_carlo}:
\bes
Z^{EPRL}_\Delta&=&
\int_{\SL(2,\C)} \prod_{e\in \Delta} dG^s_edG^t_e\,
\int_{\SU(2)}\prod_{e,f\ni e}dh_{e,f}\\
&&\prod_{f\in\Delta}
\sum_{j_f} d_{j_f}\prod_{e\in\pp f}\chi^{j_f}(h_{e,f})\,
\tr D^{(j_f,\gamma j_f )}\left(\overrightarrow{\prod}_{e\in \pp f} G^s_eh_{e,f}(G^t_e)^{-1}\right).\nn
\ees
Here, $\chi^j(h)=\tr D^j(h)$ is the character of the $\SU(2)$ group element $g$ in the representation of spin $j$, i.e the trace of the matrix representing $g$ in that representation. Similarly, $D^{(j,\gamma j )}(G)$ is the matrix representing the group element $G$ in the irreducible $\SL(2,\C)$ representation labeled by $(n=j,\rho=\gamma j)$. Finally $\gamma$ is the Immirzi parameter. There is also a proposal for a slight modification of this path integral by taking $\rho=\gamma (j+1)$ instead of $\rho=\gamma j$.
Once again, if one performs all the group integration, one will recover the spinfoam amplitude of the EPRL-FK model given as the product of local vertex amplitudes.

In both the Barrett-Crane model and the EPRL model, we have integrations over group elements living in the $\SU(2)$ subgroup of $\SL(2,\C)$. It is clearly implicit that one has chosen a particular $\SU(2)$ subgroup i.e a time normal which defines that subgroup as its stabilizer group. Actually one can change the choice of time normal and choose a different time normal for each edge $e$. This will not affect the spinfoam amplitude as defined above: the change in the time normal is a boost that can be fully re-absorbed in the $\SL(2,\C)$ group elements $G^s_e,G^t_e$. Actually, reversely, those group elements $G^{s,t}_e$ define the time normal associated to the tetrahedron dual to the edge $e$ (in both 4-simplices to which it belongs) and it is this time normal which is used to implement the simplicity constraints e.g. \cite{ls2,eprl,actionlb}.

\medskip

This concludes our overview of the construction and derivation of the spinfoam models for 4d quantum gravity based on the reformulation of general relativity as a constrained BF theory. The interested reader will find many more details on the amplitudes of these models and their properties in the references cited in the text.

\section{Group Field Theories: a Non-Perturbative Definition of Spinfoams}

{\it Group field theories} were originally introduced as auxiliary field theories for spinfoam models. Similarly to matrix models (for 2d gravity), they generate the spinfoam amplitudes as their Feynman diagrams: the Feynman diagrams of the group field theories are identified as 2-complexes, then the evaluation of those Feynman diagrams reproduces exactly the considered spinfoam amplitudes. This is based on the exact same mechanism  through which matrix models generate triangulated 2-surfaces as their Feyman diagrams. More precisely, group field theories are like tensor models, which generates space-time triangulations in dimension higher than two.

At the beginning, those group field theories were an auxiliary tools allowing to derive the spinfoam amplitudes accordingly to their local symmetries. Now, they are considered as the proper way to define any spinfoam model and the corresponding non-perturbative sum over all possible 2-complexes which defines the spinfoam amplitudes for fixed boundary data (see e.g. \cite{gft}). It also seems to be the correct framework for spinfoams in order to address the complicated questions of the renormalization and the diffeomorphism invariance of those models, although only a preliminary analysis of those issues has been done up to now.

\subsection{Matrix Model for 2d (Quantum) Gravity and 2d Spinfoams}

Let us start by reviewing the simpler framework of matrix models, and how they generate non-perturbative sums over triangulated 2-surfaces. Considering as basic variable a $N\times N$ Hermitian matrice $M$, we introduce the matrix model action:
\be
S[M]\,=\,\f N2\tr M^2 +N\f\lambda{3!}\tr M^3.
\ee
More generally, we can introduce interaction terms of arbitrary higher order:
\be
S[M]\,=\,\f N2\tr M^2 +\sum_{n\ge 3}N\f{\lambda_n}{n!}\tr M^n,
\ee
but we will focus on the simpler action with a single cubic interaction term, since this will affect our discussion.

Now, we consider the corresponding partition function:
\be
Z=\int [dM]\, e^{-S[M]}\,=\,\int [dM]\, e^{\f N2\tr M^2 +N\f\lambda{3!}\tr M^3}.
\ee
We expand this partition function, using the quadratic term as the propagator and the cubic term as the interaction vertex:
\be
Z=\int [dM]\, e^{\f N2\tr M^2}\,\sum_V\f1{V!}\left(\f{N\lambda}{3!}\right)^V\,\left(\tr M^3\right)^V.
\ee
This leads to Feynman diagrams with the interaction vertices glued together using the trivial propagator. The link with triangulated 2-surfaces is obtained by taking the topological dual: the cubic interaction becomes a triangle and Feynman diagrams consist in gluing these triangles together along their edges. Actually the Feynman diagram is the dual of the 2d triangulation and is in fact the spinfoam 2-complex itself. This trick allows to write the partition function as  a sum over all triangulated 2-surfaces:
\be
Z=\sum_{S} w(S) \lambda^V N^{F-E+V}=\sum_{g\in\N} N^{2-2g}\sum_V w_{g,V} \lambda^V= \sum_{g\in\N} w_g(\lambda)\,N^{2-2g}.
\ee
Considering a triangulated surface $S$, the number of interaction vertices $V$ is the number of triangles. The number of faces $F$ of the Feynman diagrams is the number of points of the triangulation, while the number of propagators gives the number of edges $E$. Finally $w(S)$ is the symmetry factor of the surface $S$ seen as a Feynman diagram. As it is well-known, we get a purely topological amplitude: the evaluation of a Feynman diagram, up to the obvious factor $\lambda^V$, only depends on the Euler characteristic of the corresponding triangulated surface, $\chi=F-E+V=2-2g$, where $g\in\N$ is the genus of the surface. Remember $g=0$ for a 2-spere, $g=1$ for a 2-torus and so on. The weight $w_{g,V}$ gives the number of surfaces with genus $g$ and made of $V$ triangles, up to the symmetry factors $w(S)$. Summing those weights over $V$ together with the factor $\lambda^V$ leads to the statistical factor $w_g(\lambda)$ defining the weight of the genus $g$ in the partition function.

If we were further considering higher order interaction terms, we would generate 2-surfaces made non-only from triangles but from also higher order polygons. It would only affect the definition of the statistical weight $w_g(\lambda)$ , but would not affect the topological nature of the partition function.

I will not review how to compute the partition function and correlations of the matrix model, but I simply point out that these models are completely integrable with methods based on, either the diagonalization of the matrix $M$ and the re-writing of all the amplitudes in term of its eigenvalues, or straightforward counting of the number $w_{g,V}$ of triangulations for genus $g$ and $V$ triangles.

\medskip

We have thus seen how the partition function of this simple matrix model generates a non-perturbative sum over all 2-surfaces with all possible topologies. The purpose of a group field theory  is to mimic this procedure and generate all possible (relevant) 2-complexes as its Feynman diagrams.

Before moving on to group field theories for 3d and 4d quantum gravity, I would like to define the group field theory for 2d spinfoams. This should be considered as a simple toy model, without deep physical meaning. It is merely a simple mathematical model, which allows to understand and study certain generic features of group field theories. So, taking $\SU(2)$ as gauge group, we introduce the group field $\vphi(g_1,g_2)$ on $\SU(2)^2$ and we require that it is invariant under the right diagonal $\SU(2)$-action:
\be
\vphi(g_1,g_2)=\vphi(g_1h,g_2h),\qquad\forall h\in\SU(2).
\ee
Then we define the group field theory action, similarly to the matrix model above \cite{GFT2d}:
\be
S_{2d}[\vphi]\,=\,
\int dg_1dg_2\, \vphi(g_1,g_2)\overline{\vphi}(g_1,g_2)
+\f\lambda{3!}\int dg_1dg_2dg_3\, \vphi(g_1,g_2) \vphi(g_2,g_3) \vphi(g_3,g_1).
\ee
Usually we further require that the field satisfy a reality condition, $\overline{\vphi}(g_1,g_2)=\vphi(g_2,g_1)$.

Similarly to the matrix model, Feynman diagrams of this field theory over $\SU(2)^2$ are identified as triangulated 2-surfaces, and the evaluation of these Feynman diagrams leads to (topological) spinfoam amplitudes in the connection representation \cite{GFT2d}.

As before, we can generalize this action to take into account arbitrary two-dimensional cellular complexes made of arbitrary polygons:
\be
S_{2d}[\vphi]\,=\,
\int dg_1dg_2\, \vphi(g_1,g_2)\overline{\vphi}(g_1,g_2)
+\sum_n\f{\lambda_n}{n!}\int [dg]^n\, \vphi(g_1,g_2) \vphi(g_2,g_3).. \vphi(g_n,g_1).
\ee

A final remark  about this 2d group field theory is that we can fully gauge-fix the $\SU(2)$ gauge invariance of the field $\vphi$ and express everything in term of a gauge-invariant field $\phi$:
\be
\vphi(g_1,g_2)=\phi(g_1(g_2)^{-1}).
\ee
Then the action reads \cite{GFT2d,GFTmatrix}:
\be
S[\phi]=\int dg\,\phi(g)\overline{\phi}(g)
+\sum_n\f{\lambda_n}{n!}\int [dg]^n\,\delta(\prod_{i=1}^n g_i)\,\prod_{i=1}^N \phi(g_i),
\label{GFT2d}
\ee
where the interaction terms are given by the evaluation at the identity of the convolution powers of the field $\phi$.
This reformulation of the initial 2d group field theory is particularly useful when making the link between group field theories and non-commutative field theories. In particular, the $\SU(2)$ group element $g$ is interpreted as the momentum and the factor $\delta(\prod_i g_i)$ is identified as the law of conservation of momentum \cite{GFTmatrix,GFTsym}.

\subsection{Group Field Theories for Spinfoam Models}

The matrix model logic and the construction of the 2d group field theory can be generalized at all local spinfoam models. This was finally proven in \cite{gftrr}, following techniques introduced for the Boulatov group field theory for the Ponzano-Regge model \cite{boulatov}, the Boulatov-Ooguri group field theory for 4d topological BF theory \cite{GFT_ooguri} and the DePietri-Freidel-Krasnov-Rovelli group field theory for the Barrett-Crane model \cite{fdpkr,GFTbc}. %
Now group field theories provide the final definition of spinfoam framework and any new spinfoam models is defined through its group field theory.

As its name clearly suggests, a group field theory is a field theory on a group manifold, usually defined as many copies of the relevant gauge group. Here are the basic features for a group field theory for a spinfoam model in a $n$-dimensional space-time:
\begin{itemize}

\item The group field $\vphi$ is a field on $\cG^n$, where $\cG$ is the relevant gauge group (usually $\SU(2)$ or the Lorentz group). We further require that it is invariant under a global $\cG$-action:
    \be
    \vphi(g_1,..,g_n)=\vphi(g_1g,..,g_ng),\qquad\forall g_i,g\in\cG.
    \ee
    This gauge invariance ensures that the basic excitations of the field (when decomposing it on $\cG$-representations) are $\cG$-intertwiners. Then the group field $\vphi$ appear as the basic building block of the spinfoam model: the spinfoam amplitudes are indeed all about gluing intertwiners together in an appropriate way in order to create space-time structures.

\item The action $S[\vphi]$ is made of a quadratic kinetic term defining the propagator and of (at least) one higher order term defining the interaction vertex. The propagator defines the class of field $\vphi$ that propagate, i.e the type of intertwiners from which the spinfoam is made. Historically, the usual choice of propagator was simply a projector, which was defining a ``trivial" propagation. Nevertheless, the more recent and elaborate group field theories are based on using non-trivial kinetic terms in the action, e.g. \cite{GFT_vincent}.

\item Feynman diagrams of the group field theory are defined as all the possible ways of gluing interaction vertices using the propagator. Geometrically, these Feynman diagrams are 2-complexes that are identified as dual to $n$-dimensional (pseudo-)triangulations (or more generally $n$-dimensional cellular complexes). More precisely, the group field is interpreted as the basic $(n-1)$-dimensional building block for the space triangulation, since intertwiners are the basic building block for spin network states defining the quantum geometry of the $(n-1)$-dimensional spatial hypersurface. The interaction term of the group field theory is identified to a $n$-dimensional space-time cells. In the Feynman diagrams, those $n$-dimensional space-time cells are then glued together using the field theory's propagator, thus creating a space-time triangulations.

    The typical interaction term that is used to define the group field theories are taken to be polynomials of order $n$ in $\vphi$ and represent $n$-simplices. However, we could further include in the theory any higher order interaction terms representing more complicated $n$-dimensional cells. Actually all these extra terms need to be taken into account in the effective field theory when studying the renormalization of the group field theory, or equivalently the coarse-graining of spinfoam amplitudes.

\item As we have just said, the Feynman diagram of the group field theory are mapped onto 2-complexes $\Delta$ dual to triangulations of the $n$-dimensional space-time. Then the evaluation of those Feynman diagrams define the spinfoam amplitude $\cA[\Delta]$ associated to that 2-complex. One can use directly the definition of the group field theory in term of $\vphi(g_1,..,g_n)$ and one obtains the spinfoam amplitude written in the connection representation. Or one can decompose the group field $\vphi$ in $\cG$-representations and then evaluated the Feynman diagrams, which will lead to the local spinfoam ansatz written in term of a sum over representations and interwtiners.

\item The partition function of the group field theory,
\be
Z\,=\,\int [d\vphi]\, e^{-S[\vphi]} =\int [d\vphi]\, e^{-S_{kin}[\vphi]}\,e^{-\lambda\,\cV[\vphi]}
\,=\, \sum_\Delta w(\Delta)\,\lambda^{V(\Delta)}\,\cA[\Delta],
\ee
defines the non-perturbative sum over all possible spinfoam 2-complexes through its perturbative expansion into Feynman diagrams. $S_{kin}$ is the quadratic part of the action defining the propagator; $\cV[\vphi]$ is the interaction term, which represent a $n$-simplex; $\lambda$ is the group field coupling constant, $V(\Delta)$ and $w(\Delta)$ are respectively the number of spinfoam vertices in $\Delta$ and its symmetry factor.

\item Finally, one defines the spinfoam amplitude for fixed boundary data through field insertions in the partition function, i.e non-trivial correlation functions, \cite{bubble}:
    \be
    A[\Gamma,\psi_\Gamma]
    \,=\,
    \sum_{\Delta|\pp\Delta=\Gamma} w(\Delta)\,\lambda^{V(\Delta)}\,\cA[\Delta]
    \,=\,
    \int [d\vphi]\, P_\psi[\vphi]\,e^{-S[\vphi]},
    \ee
    where $P_\psi[\vphi]$ is a polynomial functional of the group field $\vphi$ which depend on the choice of spin network state $\psi$ on the boundary graph $\Gamma$. Since $\vphi$ represents an intertwiner, the order of the polynomial $P_\psi[\vphi]$ is the number of intertwiners in $\psi$, i.e. the number of vertices in the boundary graph $\Gamma$.

\end{itemize}

We can try to summarize the various structures in the following table:

\begin{center}
\begin{tabular}{c|c|c}
4d Geometry & Spinfoam & Group Field Theory (GFT)\\ \hline
4d triangulation & 2-complex & Feynman diagram \\
4-simplex amplitude & vertex amplitude & interaction term \\
tetrahedron weight & edge amplitude & propagator \\
quantum tetrahedron & intertwiner & group field \\
boundary 3d triangulation & spin network & observable insertion
\end{tabular}
\end{center}

These are the basic features of the group field theory formalism for spinfoam models. Now, let us give explicitly some of the most used examples of group field theory.

We start with the Boulatov field theory, which generates the Ponzano-Regge amplitude for 3d quantum gravity \cite{boulatov}. The gauge group is $\SU(2)$ and the group field $\vphi$ lives on $\SU(2)^3$. It is assumed to satisfy a $\SU(2)$ gauge invariance:
\be
\vphi(g_1,g_2,g_3)=\vphi(g_1g,g_2g,g_3g),\forall g\in\SU(2).
\ee
Then the Boulatov action reads:
\bes
S_{3d}[\vphi]&=&\f12\int [dg]^3\,\vphi(g_1,g_2,g_3)\overline{\vphi}(g_1,g_2,g_3)\\
&&+\f\lambda{4!}\int[dg]^6\,\vphi(g_1,g_2,g_3)\vphi(g_3,g_4,g_5)\vphi(g_5,g_2,g_6)\vphi(g_6,g_4,g_1).\nn
\ees
The group field $\vphi$ represents a quantum triangle. Then the interaction vertex is made of four such quantum triangles and is dual to a tetrahedron. Its Fourier decomposition into $\SU(2)$ representations gives back the 6j-symbol, which is indeed the vertex amplitude for the Ponzano-Regge model. The trivial kinetic term then leads to a trivial gluing of those tetrahedra together into 3d space-time triangulations.

Finally, there exists several variations of this action according to the reality conditions imposed on the field, the exact permutations of the group elements $g_i$ in both kinetic and interaction terms and the invariance properties of the field $\vphi$ under such permutations, e.g. \cite{fdpkr,GFTsym,winston,GFT3d_more}.

This is easily generalized to the Ooguri-Boulatov group field action for 4d BF theory \cite{GFT_ooguri}. Now, we consider a invariance field on $\SU(2)^4$:
\be
\vphi(g_1,g_2,g_3,g_4)=\vphi(g_1g,g_2g,g_3g,g_4g),\forall g\in\SU(2),
\ee
and we define the action:
\bes
S_{4d}[\vphi]&=&\f12\int [dg]^4\,\vphi(g_1,g_2,g_3,g_4)\overline{\vphi}(g_1,g_2,g_3,g_4)\nn\\
&&+\f\lambda{5!}\int[dg]^{10}\,\vphi(g_1,g_2,g_3,g_4)\vphi(g_4,g_5,g_6,g_7)\vphi(g_7,g_3,g_8,g_9)\vphi(g_9,g_6,g_2,g_{10})\vphi(g_{10},g_8,g_5,g_1).\nn
\ees
The geometrical interpretation is straightforward: the field $\phi$ is a quantum tetrahedron, the quintic interaction term represents a 4-simplex, the trivial propagator gives a trivial gluing of those 4-simplices along shared tetrahedra. Expanding the Feynman amplitudes in $\SU(2)$ representations, we recover the standard 15j-symbol of the BF spinfoam amplitudes.

It is straightforward to replace the gauge group $\SU(2)$ by an arbitrary gauge group $\cG$. For instance, we can easily use the Lorentz group $\SL(2,\C)$ instead of $\SU(2)$ if we would like to recover the spinfoam amplitude for BF theory with gauge group $\SL(2,\C)$.

\medskip

Finally we give the group field theory for the Barrett-Crane model. It uses an invariant field on $\SL(2,\C)^4$. Then the action differs from the previous one by a slight modification of the kinetic term \cite{fdpkr,GFTbc}:
\be
\begin{array}{ccll}
S_{BC}[\vphi] & = & \f12\int [dG]^4 & P\vphi(G_1,G_2,G_3,G_4)\overline{P\vphi}(G_1,G_2,G_3,G_4)\\
&&+\f\lambda{5!}\int[dG]^{10} &\vphi(G_1,G_2,G_3,G_4)\vphi(G_4,G_5,G_6,G_7)\vphi(G_7,G_3,G_8,G_9) \\
&&&\vphi(G_9,G_6,G_2,G_{10})\vphi(G_{10},G_8,G_5,G_1),
\end{array}
\ee
where the map $P$ is defined through group averaging:
\be
P\vphi(G_1,G_2,G_3,G_4)=\int_{\SL(2,\C)}dG\int_{\SU(2)^4}[dh]^4\,
\vphi(G_1Gh_1,G_2Gh_2,G_3Gh_3,G_4Gh_4).
\ee
It is straightforward to check that this leads back to the Barrett-Crane spinfoam amplitudes in the connection representation \cite{GFTbc,bubble}. This modification of the propagator of the group field theory mirrors directly the imposition of the simplicity constraints on the spinfoam amplitudes.

The group field theories for the EPRL-FK spinfoam models have a slightly more complicated explicit action. The interested readers can find the explicit expressions in \cite{fk,GFT_vincent}.

\medskip

We will conclude this review of the group field theory formalism by saying that, although it is now the standard way to define spinfoam models, there has been little progress on how to truly use group field theories to extract non-trivial correlations from the spinfoam amplitudes. There have nevertheless recently been some interesting work on rigorously  defining the partition function for group field theories \cite{GFT3d_more,GFT_sum}.


\chapter{The Semi-Classical Regime and Effective Dynamics}

In the previous chapter, we have reviewed the loop quantum gravity formalism and the spinfoam framework for a path integral for quantum gravity. This provides a well-defined picture and definition of the structure of the quantum geometry at the Planck scale. The next step, once such a model of quantum gravity is explicitly defined, is to extract predictions from it and to test it against experimental data. In order to do so, we need to investigate the semi-classical regime of spinfoam models and develop the relevant tools and techniques to probe it and extract interesting physical  data.

One of the first thing to check is that we recover general relativity  in a particular regime of the theory. In particular, we would like to identify a regime with a flat space-time at large scales and small fluctuations of the geometry. This would be the first step towards bridging between the spinfoam framework and the standard perturbative expansion of general relativity as a quantum field theory around the flat space-time. This would help understanding how to use spinfoam models in practice to actually compute quantum gravity corrections to the classical gravitational dynamics.

A second direction to investigate is to study the propagation of matter fields and particles coupled to quantum gravity. One would like to understand the effect of the quantum fluctuations of the geometry onto the dynamics of matter: the goal would be to show how to recover the standard dynamics of matter propagating on a fixed non-fluctuating flat space-time, and then to be able to compute the quantum gravity correction to this matter dynamics using an appropriate perturbative expansion.

These two questions have started being addressed in the spinfoam framework through respectively two recent advances in this field:
\begin{itemize}

\item the ``spinfoam graviton propagator" where one computes correlations between geometric observables on a semi-classical state peaked on the flat space-time,

\item the coupling of matter Feynman diagrams to spinfoam amplitudes and the resulting link between spinfoam models and non-commutative field theory.

\end{itemize}

I will present these two topics in the next sections.

\section{Geometrical Correlations and the Graviton Propagator}

The goal of the proposal for computing a ``spinfoam graviton propagator" \cite{grav1} is to probe the geometry induced by the spinfoam amplitudes through calculating correlations between geometric observables. This is exactly the same as computing $n$-point correlation  functions in standard quantum field theory using the path integral formalism with Feynman diagrams and all. One of the objective is to identify appropriate observables and quantum states so that we can reconstruct from these spinfoam correlations the graviton propagator $\la h_{\alpha\beta} h_{\gamma\delta}\ra$ describing the quantum fluctuations of the gravitational field around a flat background.

The more ambitious objective of the program is, first, to show to recover classical gravity with the Newtonian law and spin-2 gravitational waves and, second, to understand how to recover the whole standard perturbative framework of quantum general relativity around the flat metric and to see how the spinfoam framework allows to address and solve (or side-step) the (na\"ive) non-renormalisability of general relativity as a quantum field theory.

\subsection{The General Framework for Spinfoam Correlations}

We start with a 3d triangulation, which will be considered as the boundary triangulation. It is dual to a graph $\Gamma$.  We define a spin network state $\psi_\Gamma$ on this boundary graph $\Gamma$, and we call it our boundary state. The corresponding spinfoam amplitude is the sum over all compatible bulk triangulations, or equivalently over all 2-complexes $\Delta$ such that $\pp\Delta=\Gamma$. Then we need to ensure too that the spinfoam data in the bulk, the representations $j$ and intertwiners $i$ are compatible with our choice of boundary state $\psi_\Gamma$. In the end, the spinfoam amplitude simply reads:
\be
\cA[\Gamma,\psi_\Gamma]=
\sum_{\Delta |\pp\Delta=\Gamma} \sum_{j_f,i_e} \sum_{j_\pp,i_\pp}
w(\Delta)\,\psi(j_\pp,i_\pp)\,\cA_\Delta [j_\pp,i_\pp,j_f,i_e],
\ee
where we have distinguished the sum over boundary representations and intertwiners, on which depends the boundary state, and the sum over representations and intertwiners in the bulk. $\cA_\Delta [j_\pp,i_\pp,j_f,i_e]$ is the spinfoam amplitude associated to the 2-complex $\Delta$ dressed with all the boundary and bulk representations and intertwiners. Finally $w(\Delta)$ is the statistical symmetry factor to the 2-complex $\Gamma$.

Let us note $\bv$ and $\bl$ respectively the vertices and links of the boundary graph $\Gamma$ with the bar to distinguish them from the bulk data. Representations $j_\bl$ live on the graph's links and intertwiners $i_\bv$ live at the graph's nodes. Geometrically, considering the bulk and boundary triangulations, the representation $j_\bl$ gives the area of the triangle $t$ dual to the link $\bl$ and the intertwiner $i_\bv$ describe the shape and volume of the tetrahedron $T$ dual ot the vertex $\bv$. The boundary representations and intertwiners are the typical geometric observables that we consider for the  ``spinfoam graviton propagator"  and that we insert in the spinfoam partition function. Indeed these defines the correlations between areas and volumes at different points on the boundary, from which we can reconstruct correlations between the various components of the metric \cite{grav1,grav2,alesci1,alesci2,alesci3}. Thus, for example focusing on correlations between the areas of two boundary triangles, we consider:
\be
\la j_{\bl_1}j_{\bl_2}\ra_\psi
\,=\,
\f1{\cA[\Gamma,\psi_\Gamma]}\,\sum_{\Delta |\pp\Delta=\Gamma} \sum_{j_f,i_e} \sum_{j_\pp,i_\pp}
w(\Delta)\,j_{\bl_1}j_{\bl_2}\,\psi(j_\pp,i_\pp)\,\cA_\Delta [j_\pp,i_\pp,j_f,i_e],
\ee
where the original amplitude $\cA[\Gamma,\psi_\Gamma]$ without observable insertions serves as the normalization factor. One can of course insert any observable $\cO(j_\pp,i_\pp)$ in this expression instead of the simple $j_{\bl_1}j_{\bl_2}$. Nevertheless, this basic choice of correlations between two area observables  is (almost) enough to fully re-construct the full graviton propagator defining the 2-point correlation functions of the metric components.

This whole framework crucially depends on the specific choice of the boundary state. We need to require two strong conditions to that boundary state $\psi_\Gamma$:
\begin{itemize}
\item We need a {\it physical state}, i.e solution to all the (Hamltonian) constraints. This translates in the spinfoam framework into straightforward normalization conditions:
    \be
    \la \psi|\psi\ra=\sum_{j_\pp,i_\pp} \bar{\psi}(j_\pp,i_\pp)\psi(j_\pp,i_\pp)=1,
    \ee
    \be
    \la \cA|\psi\ra = \cA[\Gamma,\psi_\Gamma] =\sum_{\Delta |\pp\Delta=\Gamma} \sum_{j_f,i_e} \sum_{j_\pp,i_\pp}
w(\Delta)\,\psi(j_\pp,i_\pp)\,\cA_\Delta [j_\pp,i_\pp,j_f,i_e]=1.
    \ee
    This condition has often been neglected in the correlation calculations done up to now, apart in e.g. \cite{phys1,phys2}.

\item We would like a {\it semi-classical state} peaked on a semi-classical boundary geometry and such that it induces a flat space-time metric in the bulk. An explicit Gaussian ansatz for such a semi-classical boundary state was the key proposal by Rovelli \cite{grav1}, which launched this whole spinfoam graviton programme.

\end{itemize}

Furthermore, we can add a third condition, which is often implicit in the calculations. Indeed, up to now, all the amplitudes are defined as a sum over all possible compatible 2-complexes $\Delta$. However, nobody has yet developed techniques allowing to probe analytically non-perturbative aspects of such sums. Thus, most of the research work has focused on  selecting a suitable fixed 2-complex for the bulk and computing the correlations using that bulk triangulation as background. For this approximation to be physically relevant, it leads us to a third criteria for the boundary state:
\begin{itemize}
\item We want a boundary state such that the spinfoam amplitude $\cA[\Gamma,\psi_\Gamma]$ is peaked on a specific 2-complex $\Delta$, or at least selects a family of 2-complexes that could be related to each other by some gauge transformations (discrete diffeomorphisms). This is however a criteria which is much more difficult to implement and test.

\end{itemize}

\subsection{The Practical Calculations of the Graviton Propagator}

This  framework for the spinfoam graviton propagator is based on a very simple setting. There has been a lot of research work done on this subject. Results are, up to now, both full of promise and very restricted.

On the positive side, we are able to compute systematically at leading order the spinfoam graviton propagator at large scale (for large values of the boundary areas $j_\pp$) for all the spinfoam models which have been defined. We have even developed  techniques to extract (in principle) all quantum corrections of arbitrarily higher order (interpreted as ``loop corrections"). This leads to recover the proper scaling of Newton's law for gravity, with the gravitational potential going as the inverse distance, and even the correct spin-2 tensorial structure of the graviton (correlations) for specific spinfoam models. We even understand the relation between the spinfoam path integral and Regge calculus at large scale. The short scale behavior has also been investigated. It appears that the graviton propagator is regularized (as expected) by quantum gravity effects and that we have the emergence of a dynamical minimal length scale close to the Planck scale. All this has been tested analytically and numerically.

On the negative side, the actual setting for practical calculations of the spinfoam graviton correlations has been much too simple up to now. These basic calculations were done mainly for a single 4-simplex, which is indeed the simplest space-time triangulation. They typically do not involve summing over bulk internal associated to internal spinfoam vertices.
Thus, these calculations don't allow to truly test the quantum gravity dynamics defined by the spinfoam models and the gluing of 4-simplices (``space-time atoms") used to construct the amplitudes. They should be considered as kinematical checks. It thus remains a challenge to go beyond the single 4-simplex and work with refined space-time triangulations, which would allow local fluctuations of the curvature in the bulk.

Here is nevertheless a (almost-exhaustive) list of the works done on the programme of the spinfoam graviton propagator:

\begin{itemize}

\item Definition of the framework \cite{grav1,grav2}.

\item Analytical study of the asymptotic ansatz for the spinfoam vertex amplitude in order to recover at leading the correct tensorial structure for the graviton propagator \cite{alesci1,alesci2,alesci3}.

\item Group integrals techniques to compute explicitly analytically the graviton propagator for the Barrett-Crane model (generalizable to arbitrary spinfoam models expressed in the connection representation) \cite{grav3}.

\item Numerical investigations of the behavior of the graviton propagator for the Barrett-Crane model, both for the large scale and short scale, both at leading order and at next-to-leading order (first order quantum gravity corrections) \cite{num1,num2}

\item Calculations of the asymptotics of the graviton propagator for the EPRL-FK spinfoam model \cite{gravEPRL}.

\item  Definition of a 3d toy model using the Ponzano-Regge model \cite{grav3d1}, numerical investigations and development of the tools to compute the full expansion of the correlations and solve the model analytically \cite{grav3d2,grav3d3}.

\item Study of the propagation of coherent wave-packets of geometry within a 4-simplex \cite{grav_coh}.

\item Analytical and numerical study of the asymptotics of the spinfoam vertex amplitude relevant to the calculations of the large scale behavior of the graviton propagator, in 3d \cite{asympt3d,asympt3d2} and in 4d for both the Barrett-Crane model \cite{asymptBC} and the EPRL vertex amplitude \cite{asymptEPRLl,asymptEPRLm,asymptEPRLn}.

\item Discussion of the potential use of the recursion relations satisfied by the spinfoam vertex amplitudes to the computation of the graviton correlations and to derivation of Ward-Takahashi identities for spinfoam amplitudes \cite{asympt3d2,recursion}.

\item Tentative calculations of the 3-point correlation functions \cite{grav3pts}.

\end{itemize}


What is very nice about this framework is that it provides a physical interpretation to the correlations computed using spinfoam models and in particular shows how to recover the classical Newton's law for gravity from our complicated and intricate model for a quantum gravity path integral. Moreover, we can actually compute analytically these correlations, plot them numerically, check that everything is consistent, and see explicit the first elements of the spinfoam dynamics with our own eyes.

However, progress in this direction is completely coupled with necessary progress that needs to be done on the coarse-graining and renormalization of spinfoam models. Indeed, we need to be able both to repeat the same graviton correlation computations for more refined and complex bulk triangulations and to say something about the non-perturbative sum over all 2-complexes. The main hope for this is put in exploiting the group field theory formalism and studying its renormalization as a quantum field theory.

\section{From Spinfoam Amplitudes to Non-Commutative Field Theory}

Besides looking at the quantum gravity corrections to the gravitational interaction, another way to probe the semi-classical regime of quantum gravity and extract potential experimental predictions is to study the dynamics of matter fields coupled to quantum geometry. The strategy is to integrate out the (quantum) fluctuations of the gravitational field and determine the resulting corrections to the classical/quantum field theory describing the standard propagation of matter.

One of the expectations is that such an analysis leads to non-commutative field theories, which allow to take into account the universality of a length or mass scale, such as the Planck scale, by quantum-deforming the Poincar\'e symmetry. This leads for example to deformed/doubly special relativity based on the $\kappa$-deformed Poincar\'e group. The effective deformation of the flat space-time is supposed to take into account the integrated fluctuations of the gravitational field, which blur the notion of space-time points at the microscopic level and thus induces an effective non-commutativity of the space-time coordinates.

This is exactly what is realized in the spinfoam program, at least when looking at the propagation of particles coupled to 3d quantum gravity. Moreover, we can see the link between spinfoam models and effective non-commutative quantum field theories directly at the level of group field theories: indeed group field theories themselves turn out to be genuine non-commutative field theories with quantum-deformed symmetries. These arguments have started to be generalized to  4d quantum gravity and it looks like a very promising direction of research.

\subsection{From Spinfoam Amplitudes to Feynman Diagrams}

We start by looking at the case of 3d Euclidean quantum gravity and the corresponding Ponzano-Regge spinfoam model. Considering a particular space-time triangulation $\Delta$, the spinfoam path integral reads:
\be
Z_\Delta=\int_{\SU(2)} \prod_t dg_t\,\prod_e \delta\left(\overrightarrow{\prod_{t\ni e}} g_t\right),
\ee
where the group elements $g_t\in\SU(2)$  are associated to every triangle (dual to spinfoam edges) and the closed holonomies $G_e\,\equiv\,\overrightarrow{\prod t\ni e} g_t$ are computed around every edge of the triangulation. In this picture, particles are included as topological defects inducing non-trivial  holonomies around the particles (see e.g. \cite{pr1,pr3} and references therein). Then one can write spinfoam amplitudes coupled to particles. For instance, inserting  particles along the edges of a graph $\Gamma$ in the triangulation, we get:
\be
Z_{\Delta,\Gamma}=\int_{\SU(2)} \prod_t dg_t
\,\prod_{e\in\Gamma} F_{s_e,m_e}(G_e)
\,\prod_{e\notin\Gamma} \delta\left(G_e\right),
\ee
where $s_e$ and $m_e$ are respectively the spin and mass of the particle living on the edge $e$. The functions $F_{s_e,m_e}(G_e)$ are the relevant propagator for the particle, which can be taken either on-shell (projection on the mass-shell condition) or off-shell (Feynman propagator). For more details, the interested reader can refer to \cite{pr1,pr3}.

The key result derived in \cite{pr3,pr3bis} is that these amplitudes $Z_{\Delta,\Gamma}$ are actually evaluations of Feynman diagrams of non-commutative field theories. More precisely, let us fix all the mass to be equal, $m_e=m$, and let us consider only scalar particles. Then, as long as the bulk triangulation $\Delta$ has the trivial topology of a 3-sphere, then all the group elements $g_t$ can be intergrated upon until we are left only with an integration over the holonomies around each particle in $\Gamma$. Looking at this final expression, one realizes that it is exactly the evaluation of the Feynman diagram $\Gamma$ for a non-commutative field theory:
\be
Z_{\Delta,\Gamma}=I_\Gamma,
\label{equal}
\ee
where $I_\Gamma$ is the Feynman diagram (for the graph $\Gamma$) of a scalar field theory on $\SU(2)$ with action:
\be
S_{eff}[\phi]\,=\,\f12\int [dg] \overline{\phi}(g)\cK_m(g)\phi(g)
+\sum_{n\ge 3}\f{\lambda_n}{n!}\int [dg]^n\,\delta(\prod_i^n g_i)\,\prod_i \phi(g_i),
\label{Seff}
\ee
where the kinetic term $\cK_m(g)$ is the inverse of the propagator $F(s=0,m)$.m field theory

This calculation can of course be adapted to particles with spin and to non-trivial space-time topologies \cite{pr3}. It can also be generalized to arbitrary gauge groups other than $\SU(2)$. For example, it can be applied to standard abelian groups such as $\R^3$ or $\R^4$, in which case a side-product of this construction is that we can write all Feynman diagrams of standard quantum field theory (e.g. the standard model for particle physics) as spinfoam amplitudes \cite{pr3,baratin}. This is a beautiful side-product which shows the universality and generality of spinfoam models in theoretical physics, since all quantum field theory can be reformulated as spinfoam amplitudes.

The actual consequences of the previous procedure for spinfoam models are very interesting:
\begin{itemize}

\item The effective action is invariant under the quantum double DSU$(2)$, which can be seen as the $\kappa$-deformation of the Poincar\'e group $\ISO(3)$ for the flat Euclidean  3d space-time. The momentum space is the group manifold $\SU(2)$ and the law of conservation of momenta is imposed by the $\delta$-distributions $\delta(\prod_i^n g_i)$. The coordinate space is reconstructed by duality from $\SU(2)$ (by using a group Fourier transform), with a $\star$-product dual to the group multiplication on $\SU(2)$. The resulting space is non-commutative, since the momentum space is curved, and the $\star$-product is non-abelian. We finally check that we can not localize space-time exactly on that non-commutative space and that the minimal length is given by the Planck scale. The reader can find all details in \cite{pr3bis} and subsequent works e.g. \cite{GFTmatrix,majid,karim}.

\item This non-commutative field theory is truly the effective field theory describing the dynamics of the matter field after having integrating over all the degrees of freedom of the gravitational field. It is thus a quantum gravity effect.

\item This effective action $S_{eff}$ has a strong resemblance with the 2d group field theory $S_{2d}$ defined earlier in eqn.\Ref{GFT2d}, which suggests that there might be a deeper link between these effective non-commutative field theories and group field theories.

\end{itemize}

The seminal work \cite{pr3,pr3bis} was done in the context of Euclidean 3d gravity. Going to Lorentzian 3d gravity is straightforward. Generalizing to 4d quantum gravity is, on the other hand, much more complicated. There has been some work in the framework if the MacDowell-Mansouri reformulation of general relativity as a $\SO(4,1)$-gauge field theory  \cite{mcdm} to understand particle insertions as topological defects \cite{defect4d}. But this has not yet lead to a definitive conclusion despite promising works hinting towards an effective non-commutative field theory based on the $\kappa$-deformed Poincar\'e group \cite{dsr4d,matter4d}.

The advantage of this line of logic to probe the semi-classical limit of spinfoam models is that it leads to actual corrections to matter dynamics, which can be really used for experiments and actual quantum gravity phenomenology.

\subsection{The Interplay between Group Field Theory and NC Field Theory}

As we have pointed out in the previous section, the effective non-commutative field theory for matter field coupled to 3d quantum geometry looks almost the same as the 2d group field theory. More precisely, comparing the actions \Ref{GFT2d} and \Ref{Seff}, the two theories are based on the same field $\phi(g)$ on $\SU(2)$, the interaction terms are the same, and finally the only thing  that differs is that the kinematical term of the non-commutative field theory has a non-trivial propagator.

This clearly suggests two things:
\begin{enumerate}

\item  There might be a more direct way to derive the effective non-commutative field theory \Ref{Seff} from spinfoam models using the group field theory formalism.

\item The 2d group field theory is actually a non-commutative field theory with a deformed Poincar\'e symmetry. And maybe more generally, group field theories for 3d and 4d spinfoam models are also genuine non-commutative field theories with quantum-deformed Poincar\'e-like symmetries

\end{enumerate}

These two natural ideas actually turned out to be both true. Let us start by the first idea, about deriving more directly the  effective non-commutative field theory from group field theories. Let us look again at the equality \Ref{equal} between spinfoam amplitudes (coupled to particles) and Feynman diagram evaluations for the non-commutative field theory. Since spinfoam amplitudes are actually derived as Feynman diagrams of a group field theory, it seems natural to make a link directly between the group field theory and the effective non-commutative field theory. Indeed two field theories with equal Feynman diagrams should be in principle the same field theory. More precisely, we would like to close the box diagram:

\begin{center}
\begin{tabular}{ccc}
group field theory & $\longrightarrow$ & spinfoam amplitudes \\
${\mathbf \downarrow}?$ & \quad & $\downarrow$\\
NC field theory & $\longrightarrow$ & Feynman diagrams for NC matter
\end{tabular}
\end{center}

\noindent
with the acronym NC stands for ``non-commutative". The missing arrow was identified in \cite{winston}, which provides the direct path from the 3d group field theory for the Ponzano-Regge model down to the effective non-commutative field theory. One looks at certain 2d variations of the group field $\vphi$ around some non-trivial classical solution to the group field equation and one recovers the effective field theory (see also \cite{GFTmatrix,GFTsym}), explicitly:
\be
\vphi(g_1,g_2,g_3)=\vphi_0(g_1,g_2,g_3)+\phi(g_1g_3^{-1}),
\ee
\be
S_{eff}[\phi]\,=\,S_{3d}[\vphi_0+\phi]-S_{3d}[\vphi_0],
\ee
where $\phi(g_1g_3^{-1})$ are the 2d variations of the 3d group field $\vphi(g_1,g_2,g_3)$. The classical solution is taken of the type $\vphi_0(g_1,g_2,g_3)\propto\int dh \delta(g_1h)f(g_2h)\delta(g_3h)$ where $f$ is an arbitrary function on $\SU(2)$. The kinetic term $\cK(g)$ of the effective field theory is explicitly and entirely determined by the choice of $f$.

Not only this shows that we can go directly from the group field theory to the effective non-commutative field theory at the classical level, but it also shows that the 3d group field theory contains a physical sector with excitations which we can identify as matter degrees of freedom. This strategy can be applied to 4d spinfoam models, and we can derive from the 4d group field theory an effective non-commutative field theory invariant under the $\kappa$-deformed Poincar\'e group \cite{matter4d}, but there are still a lot of issues and open questions to solve in that context.

\medskip

The second issue to investigate is the direct interpretation of group field theories as non-commutative field theories. Similarly to the effective non-commutative field theory, we can consider that group elements are the momentum variable and take the Fourier transform in order to recover the field in the ``true" coordinate space. This works and allows to write group field theories explicitly as non-commutative field theories with quantum-deformed symmetries \cite{GFTsym,NCGFT}. Although it is very interesting and essential to realize the symmetries of the group field theories are to be understood as quantum groups, this creates more questions  than it solves, among which: what is the actual full symmetry of group field theories? what are the consequence of the quantum-deformed symmetry for the braiding and statistic of the spin? how does it affect the path integral (measure)? This are questions that have to be addressed  in order to fully understand the group field theory formalism for spinfoam models.


\chapter{What's next for Spinfoams?}

The spinfoam framework for quantum gravity is a developing field. There has been many techniques developed to study the spinfoam amplitudes but there is still a lot of issues to clear up and understand, much more than what has already been done. Here is my personal list of principal questions on spinfoams, which is actually also my research project on this field:

\begin{itemize}

\item We have to understand the renormalization of spinfoam models. For this, we need to develop a proper framework with the appropriate tools to study the coarse-graining of spinfoam amplitudes. This is fundamental in order to truly define the continuum limit of spinfoam models and their semi-classical regime.

    We should of course investigate the coarse-graining of the current proposal of spinfoam amplitudes, such as the Barrett-Crane model or the EPRL-FK models. But beyond this, we need to identify a family of spinfoam models, which would be stable under coarse-graining. We need to understand what are the possible counter-terms and effective interaction terms to the spinfoam amplitudes. We have to understand what are the physical relevant coupling constants, such as it has been done for standard quantum field theory.

    An interesting framework in which to study these issues is the definition of spinfoam amplitudes as path integral for discrete (BF) Lagrangians. We should find out what interaction terms arise under coarse-graining and what are the corresponding quantum gravity corrections to the initial Lagrangian. Then taking the continuum limit, we would see what kind of corrections to general relativity they induce.

\item There is a lot more to understand on the interplay between spinfoam models and  non-commutative field theories. This relationship works both ways. Spinfoam models and group field theory seems to provide a nice physical framework for non-commutative field theories, in particular to understand the physical interpretation and origin of the non-commutativity of the geometry. Also we can use tools from  non-commutative quantum field theory to study the properties of the spinfoam path integral. Finally, extracting consistently from spinfoam models  non-commutative field theories  describing the effective dynamics of matter fields seems to be the fastest way to deriving phenomenological predictions from the spinfoam framework for quantum gravity.

\item We need to understand the symmetries of the spinfoam amplitudes, more particularly their invariance under discrete equivalents of space-time diffeomorphisms. This is essential both physically and mathematically. Indeed, it is crucial to understand the diffeomorphism invariance of spinfoam models and the underlying physical principle(s) behind it. It will also help understanding how to recover the usual invariance under diffeomorphisms of general relativity in the semi-classical regime. Mathematically, it is necessary to identify the symmetries of the amplitudes in order to define rigorously the path integral and the sum over all spinfoam 2-complexes. Indeed, we will need to gauge-fix the diffeomorphisms in order to get finite transition amplitudes and correlations.

    A promising direction of research are the recursion relations satisfied by the spinfoam vertex amplitudes \cite{recursion}. They are clearly related to the topological invariance and to the Hamiltonian constraint for the BF spinfoam amplitude. Hopefully, we can use them in a similar way on EPRL-FK models or other spinfoam models to understand the diffeomorphism invariance and the role of the Hamiltonian constraint for spinfoam quantum gravity.

\end{itemize}


\part{Selected Publications}



In this part, I present my original contributions to research on the spinfoam framework, through a series of published papers. This is organized along four main guidelines. First, I present papers dealing with the definition of new mathematical tools for spinfoam models, among which the construction of coherent intertwiners and the study of the asymptotics of spinfoam amplitudes. In a second chapter, I discuss papers relating to the foundations of the spinfoam framework: the construction of new spinfoam models and the investigation of their geometric meaning and algebraic properties. Then the third chapter tackles the graviton  propagator in spinfoam models and I present both analytical work on the asymptotical expansion of the graviton propagators and numerical simulations confirming the expected behavior at macroscopic and microscopic scales. Finally the fourth and final chapter consists of works on the derivation of effective field theory for the matter field dynamics from spinfoam amplitudes and on the (deep) relation between, on one side, spinfoams  and group field theory and, on the other side, non-commutative field theories.

The paper titles are directly linked to the arXiv database.

\chapter{Mathematical Tools for Spinfoams}

\section{Constructing Coherent Intertwiners and Semi-Classical States}

The following paper \href{http://arxiv.org/abs/0705.0674}{``{\it A new spinfoam vertex for quantum gravity}"} as published as in 2008 in Physical  Review D76.  This work, done in collaboration with Simone Speziale, introduced the mathematical tools of coherent intertwiners. These coherent intertwiners have very interesting semi-classical properties and allow to peak intertwiners on specific classical polyhedra. This allows for a clearer geometrical interpretation of intertwiners, spin network states and also spinfoam structures. This new tool has, since then, become a standard piece in the spinfoam machinery to define new models (see for instance \cite{fk,ls2,actionfc}) and study their semi-classical asymptotical behavior (see e.g. \cite{asymptEPRLl,asymptEPRLn}).

This paper was also one of the first works advocating that the simplicity constraints have to be imposed in a weaker sense on the spinfoam amplitudes and proposing an explicit scheme to realize this. All the subsequent EPRL-FK spinfoam models, widely accepted as the best current spinfoam model for quantum gravity, are based on this observation.




\section{Asymptotics of Spinfoam Amplitudes}

Studying the asymptotics of the spinfoam vertex amplitude is a necessary step to understand and probe the large scale structure of the space-time induced by spinfoam models. It allows to make an explicit link between spinfoams and the Regge calculus for discretized general relativity. It is also an essential ingredient of the calculations of geometrical correlations and graviton propagator from spinfoam amplitudes.

In the following two papers, co-written with the PhD student Ma\"it\'e Dupuis, respectively entitled \href{http://arxiv.org/abs/0905.4188}{``{\it Pushing Further the Asymptotics of the 6j-symbol}"} and \href{http://arxiv.org/abs/0910.2425}{``{\it The 6j-symbol: Recursion, Correlations and Asymptotics}"}  and respectively published in Physical Review D in 2009 and in Classical and Quantum Gravity in 2010, we investigate the asymptotic behavior of the 6j-symbol at large spins. The 6j-symbol is the spinfoam vertex amplitude for the Ponzano-Regge spinfoam model for 3d quantum gravity. It should be considered as toy model for vertex amplitude of 4d spinfoam models. Here, we review various ways of extract these asymptotics, through group integral, sum over representations and recursion relations. Furthermore we show how to consistently extract the quantum corrections to the asymptotics order by order in a systematic way.



\chapter{Spinfoams: Foundations \& New Models}

\section{Constructing New Models}

For a long time, the Barrett-Crane model was the only explicit spinfoam models for 4d quantum gravity. It slowly appeared that it had many shortcomings and it became necessary to define new spinfoam models which would have a better semi-classical behavior and whose geometrical interpretation would be more transparent. This lead to the introduction of the EPRL-FK spinfoam models, which represent the state-of-the-art of the spinfoam framework.

The first of the following two papers, entitled \href{http://arxiv.org/abs/0708.1915}{``{\it Consistently Solving the Simplicity Constraints for Spinfoam Quantum Gravity}"} and published by Europhysics Letters in 2008,  was written in collaboration with Simone Speziale as a follow-up of the paper ``{\it A new spinfoam vertex for quantum gravity}" already presented previously  in chapter \ref{newvertex}. We apply the coherent intertwiner techniques to re-derive the EPR model, later generalized into the EPRL class of spinfoam models, and proposed an alternative model, later identified as part of the FK family of spinfoam models.

The second paper, written in collaboration with the quantum gravity team of Carlo Rovelli in Marseille, is actually the paper which defines the EPRL spinfoam models, now the current spinfoam model used in almost all spinfoam calculations. It is called \href{http://arxiv.org/abs/0711.0146}{``{\it LQG vertex with finite Immirzi parameter}"} and was published by Nuclear Physics B in 2008.



\section{Linking the Path Integral to the Canonical Framework}

This work, \href{http://arxiv.org/abs/1008.4093}{``{\it Lifting SU(2) Spin Networks to Projected Spin Networks}"}, was realized with my PhD student Ma\"it\'e Dupuis. It shows rigorously that the space of boundary states of spinfoam models, identified as projected spin network, is isomorphic to the space of spin network states of the canonical loop quantum gravity framework. This isomorphism is crucial to truly interpreting spinfoam amplitudes as transition amplitudes for spin networks, and thus claiming that spinfoam models genuinely implement the dynamics for loop quantum gravity.

In this paper, we introduce the explicit isomorphism, investigate the ambiguities in its definition and study its mathematical properties (in particular, its unitarity).
It was published in 2010 by Physics Review D.


\section{Discretization and Action Principle for Spinfoams}

Spinfoam amplitudes can be constructed and derived from various starting point and using different procedures. One particularly rigorous point of view is to derive these spinfoam amplitudes as the path integral of discrete BF Lagrangian defined on space-time triangulations. This is exactly the perspective developed in \href{http://arxiv.org/abs/0812.3456}{``{\it A Lagrangian approach to the Barrett-Crane spin foam model}"}, written with my PhD student Valentin Bonzom and published by Physical Review D in 2009.

We define discrete Lagrangian of the BF type on space-time triangulations and compute the resulting path integral explicitly. In particular, we discuss extensively how to impose the constraints on the $B$-variables in the discrete setting and the ambiguities in defining the path integral measure. We finally introduce discrete Lagrangian for both the Barrett-Crane model and the EPRL-FK models. We further propose how to modify those path integrals either by changing the constraints or changing the path integral measure.
We also discuss the geometrical interpretation of the resulting spinfoam models.


\section{Understanding the Symmetries of Spinfoam Models}

One key issue that remains to be understood is the symmetries of spinfoam amplitudes and how are diffeomorphisms represented in this discrete setting. In this work, realized in collaboration with Valentin Bonzom and Simone Speziale, we propose to use the tool of recursion relations to explore this topic. Indeed, recursion relations are, on the one hand, very useful tools to compute numerically spinfoam amplitudes and also to extract analytically their asymptotics, but they turn out, on the other hand, to be intimately related to the topological invariance of BF theory. Here, we show this relation explicitly, i.e. how to derive recursion relations from topological invariance and, vice-versa, how recursion relations implements the quantum Hamiltonian constraint(s) generating the topological invariance. We finally provide a new way to derive recursion relation for the Barrett-Crane model. We hope that this framework can be applied to the EPRL-FK models, and other to-be-proposed spinfoam models, and will help understand how discrete diffeomorphisms are implemented in them.

This paper \href{http://arxiv.org/abs/0911.2204}{``{\it Recurrence relations for spin foam vertices}"} was published by Classical Quantum Gravity in 2010.


\chapter{Graviton Propagator \`a la Spinfoam}

\section{Analytical Methods: Group Integrals and Asymptotics}

The two following papers develop analytical techniques to compute and study the asymptotical behavior of the graviton propagator in spinfoam models.

The first one, \href{http://arxiv.org/abs/gr-qc/0608131}{``{\it Group Integral Techniques for the Spinfoam Graviton Propagator}"} published by the Journal of High Energy Physics (JHEP) in 2006, explore the use of group integrals to compute the graviton propagator. It leads to an analytical derivation of the asymptotics of the graviton propagator for the Barrett-Crane model.

The second paper explores the 3d toy model for quantum gravity amplitudes introduced earlier by Simone Speziale \cite{grav3d1}. We describe the structure of the asymptotical series of the geometrical correlations. We show how the leading order fits the expectation from 3d Regge calculus and we explain the physical origin of each of the quantum corrections. Finally we discuss the effect of the path integral measure on the amplitude of those higher order quantum corrections. We further provide numerical simulations to check each of our analytical claims.
This paper, written in collaboration with Simone Speziale and Joshua Willis, is entitled \href{http://arxiv.org/abs/gr-qc/0605123}{``{\it Towards the graviton from spinfoams: higher order corrections in the 3d toy model}"} and was published by Physical Review D in 2007. It was later followed by \cite{grav3d3}, which I wrote with Valentin Bonzom, Matteo Smerlak and Simone Speziale, which show how to consistently compute all of the higher order corrections to the leading order asymptotic behavior of the geometrical correlations of this 3d toy model.

\section{Numerical Results: Large Scale \& Short Scale Behavior}

These two papers, \href{http://arxiv.org/abs/0710.0617}{``{\it Numerical evidence of regularized correlations in spin foam gravity }"} and \href{http://arxiv.org/abs/0908.4476}{``{\it Sub-leading asymptotic behaviour of area correlations in the Barrett-Crane model }"},  present numerical simulations of the graviton propagator for the Barrett-Crane spinfoam model for 4d quantum gravity. They confirm the analytical calculations of the leading order and first order corrections and they explore the short scale behavior showing that  the propagator is indeed regularized and that a minimal length scale appears dynamically.
The first one appeared in Physics Letters B in 2008 and the latter one was published by Classical and Quantum Gravity later in 2010.

\chapter{Extracting Matter Dynamics from Spinfoam Amplitudes}

This last chapter present a series of three papers on the topic of deriving effective field theory describing the dynamics of matter evolving coupled to quantum gravity, and exploring the resulting interplay between group field theories for spinfoam models and non-commutative field theories.

The first paper, \href{http://arxiv.org/abs/hep-th/0512113}{``{\it  3d Quantum Gravity and Effective Non-Commutative Quantum Field Theory}"}, written with Laurent Freidel, was published by Physics Review Letters in 2006. It is a shorter version of \cite{pr3} focusing on the actual derivation of the effective non-commutative field theory describing the dynamics of a scalar field coupled to 3d quantum gravity. We write the spinfoam amplitudes for particles coupled to the 3d quantum geometry and show that they are equal to the Feynman diagrams of a  effective non-commutative field theory. We then investigate the properties  of this field theory. In particular, we introduce the resulting $\star$-product on the non-commutative $\R^3$ space. This $\star$-product has been thoroughly studied in subsequent works e.g. \cite{majid,karim}, and is now part of the standard tools of the spinfoam framework. It has turned out particular useful to study the structure and symmetries of group field theories \cite{GFTsym, NCGFT} or to construct discrete actions for spinfoam amplitudes \cite{actionlb} or even to revisit the loop quantum gravity framework \cite{NCLQG}.

The second paper presents work realized in collaboration with Winston Fairbairn. Entitled \href{http://arxiv.org/abs/gr-qc/0702125}{``{\it 3d Spinfoam Quantum Gravity: Matter as a Phase of the Group Field Theory}"}, it was published by Classical and Quantum Gravity in 2007. This is the paper that showed how to derive directly the effective non-commutative field theory from the group field theory directly at the level of the classical action. This opened the door to the interpretation of group field theories as non-commutative field theories. Moreover, this paper proposed the novel procedure of introducing a non-trivial geometrical background in the totally background-independent framework of group field theories by looking at perturbations around non-trivial classical solutions of the group field theory.

Finally, the last paper \href{http://arxiv.org/abs/0811.1462}{``{\it Matrix Models as Non-commutative Field Theories on R${}^3$}"} was published by Classical and Quantum Gravity in 2009. It investigates the properties of the $\star$-product on the non-commutative $\R^3$ space introduced in the earlier paper ``{\it  3d Quantum Gravity and Effective Non-Commutative Quantum Field Theory}", and uses it to show the interplay between group field theories, non-commutative field theories and matrix models.





\chapter*{Acknowledgments}
\addcontentsline{toc}{chapter}{Acknowledgments}

Thanks to everybody who has collaborated and worked and discussed with me about physics during all these years! Thanks to those special ones who made my time in research more fun! And thanks to everyone who managed to read this thesis up to here!







\end{document}